\documentclass[journal]{IEEEtran}

\ifCLASSINFOpdf
   \usepackage[pdftex]{graphicx}
\else

\fi

\usepackage[english]{babel}
\usepackage[hidelinks]{hyperref}
\usepackage{graphicx}
\usepackage{algorithm}
\usepackage{algorithmic}
\usepackage{multirow}
\usepackage{subfigure}
\usepackage{color}
\usepackage{amsmath}
\usepackage{epstopdf}
\usepackage{url}

\newcommand{\algorithmicJUAN}{\textbf{Method:}}
\newcommand{\JUAN}{\item[\algorithmicJUAN]}
\begin{document}

\title{Paper evolution graph: Multi-view structural retrieval for academic literature}

\author{Danping~Liao,
 Yuntao~Qian~\IEEEmembership{Member,~IEEE,}

\thanks{D. Liao and Y. Qian are with the Institute of Artificial Intelligence, College of Computer Science, Zhejiang University, Hangzhou 310027,
P.R. China. Corresponding author: Y. Qian (ytqian@zju.edu.cn)}
}

{~~}
\maketitle

\begin{abstract}
Academic literature retrieval is concerned with the selection of papers that are most likely to match a user's information needs.
Most of the retrieval systems are limited to list-output models, in which the retrieval results are isolated from each other.
In this work, we aim to uncover the relationships of the retrieval results and propose a method for building structural retrieval results for academic literatures, which we call a paper evolution graph (PEG).
A PEG describes the evolution of the diverse aspects of input queries through several evolution chains of papers.
By utilizing the author, citation and content information, PEGs can uncover the various underlying relationships among the papers and present the evolution of articles from multiple viewpoints.
Our system supports three types of input queries: keyword, single-paper and two-paper queries.
The construction of a PEG mainly consists of three steps.
First, the papers are soft-clustered into communities via metagraph factorization during which the topic distribution of each paper is obtained.
Second, topically cohesive evolution chains are extracted from the communities that are relevant to the query. Each chain focuses on one aspect of the query.
Finally, the extracted chains are combined to generate a PEG, which fully covers all the topics of the query.
The experimental results on a real-world dataset demonstrate that the proposed method is able to construct meaningful PEGs.
\end{abstract}


\IEEEpeerreviewmaketitle

\section{Introduction}
\label{intro}
Where did the idea of this paper come from?
Are there any improved methods to do this?
These are the questions beginners try to find answers when faced with an unfamiliar research territory.
However, as the academic literatures become ubiquitous, the problem of information overload has arisen.
Users find an overwhelming number of publications that match their search queries but they can still be confused about where to start.
For this reason, there is a growing need for techniques that can present the retrieved papers in a meaningful and effective way.

The existing academic literature retrieval systems such as Google Scholar, Scopus and Web of Science
play an important role in retrieving articles of interest.
Applying advanced ranking algorithms, these systems are able to return articles that are most likely to match users' queries.
However, although these systems are effective in retrieving the relevant papers, the returned articles are displayed in a listed and isolated way. In other words, the underlying relationships between the retrieved articles remain unknown to the users.
It is still a problem for the users to make a reading plan that guides them on to what to read first and next.


Some systems move beyond the list-output models and provide structural retrieval results.
For example, Web of Science creates a citation map for each query-paper based on its forward and backward citation relationships.
However, when the query-paper cites or is cited by a large number of papers, the papers are squeezed together such that it is hard to discern the papers.
In addition, only the citation connection between the papers is elicited: there is no content/topic information presented.
It still takes great effort for users to locate their papers of interest within such a big map.
\begin{figure}[t]
  \centering
     {\includegraphics[width=0.8\linewidth]{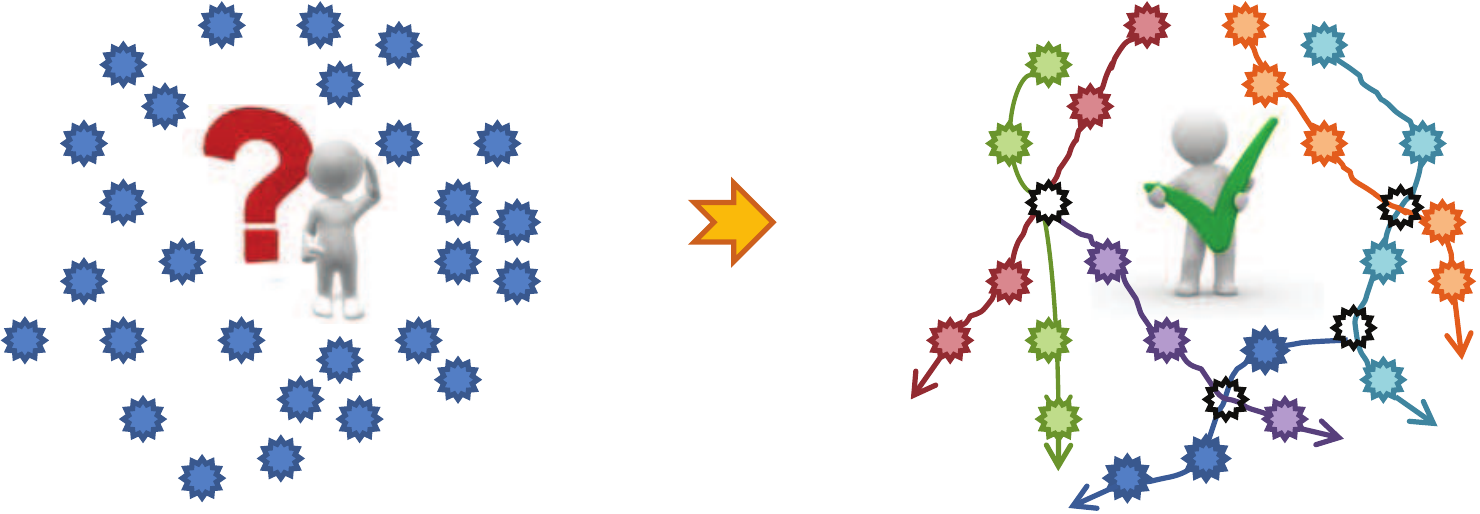}}
\caption{Finding paths between pieces of messy information.}\label{fig:Finding paths among messy information}
\end{figure}

\begin{figure}[t]
  \centering
     {\label{a}\includegraphics[width=0.8\linewidth]{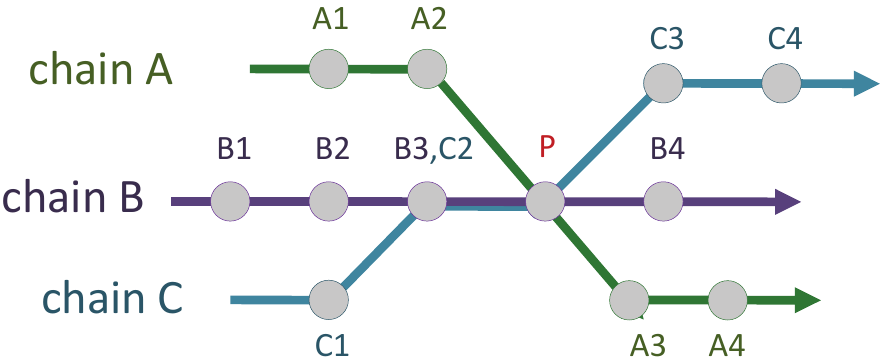}}
\caption{A simplified example of a PEG.}\label{fig:PEG example}
\end{figure}

In this paper, we present the retrieved results in a way that explicitly shows the evolutionary relationship between the papers.
As shown in Fig.~\ref{fig:Finding paths among messy information},
our system aims to string the retrieved articles together in an evolutional way and combine the strings to form a graph, which we call a paper evolution graph (PEG).
Fig.~\ref{fig:PEG example} shows a simplified PEG.
As can be seen, a PEG is a combination of several evolution chains depicted by different colors.
Each evolution chain consists of a set of topically cohesive papers.
Different chains focus on the evolution of the different topics relevant to the query.
The common nodes of different chains reveal the intersection of different topics.
For example, the PEG in Fig.~\ref{fig:PEG example} is generated based on an  input query-paper $\textit{P}$, which we assume applies three techniques to achieve a final goal.
This PEG consists of three evolution chains describing the three technical routes that $\textit{P}$ involves.

A PEG allows users to browse the retrieved papers at a holistic level and navigate the overall aspects of the query.
To fully uncover the relationships between academic articles, our system utilizes the content, author and  citation  information to discover latent relationships from multiple viewpoints between papers.
This allows our system to incorporate user preference to generate PEGs focusing on different types of coherence.
For example, a PEG emphasizing author coherence is likely to consist of chains of articles published by the same authors, while a PEG emphasizing citation coherence tends to consist of articles that have citation relationships.

The process of building a PEG can be summarized in three steps.
First, to fully cover the topics of a query-paper in a PEG, we obtain the topic distribution of the papers in the dataset by multi-relational factorization of the metagraph of our system.
After this step, articles in the dataset are soft-clustered into communities based on their topic distribution.
Second, from each community that relates to the query, the most cohesive chain is extracted based on a proposed criterion for topic coherence.
Finally, the extracted chains are combined to form a PEG.

To satisfy different user requirements, our system supports three types of queries: keyword, single-paper and two-paper queries.

\emph{Search by keyword:}
Beginners who are new to an academic domain are always curious about the overall development of that domain. For example, a student who is new to ``deep learning'' would first search the words ``deep learning'' and then read a most classical article as suggested.
Yet, what is the reader to do next? One option is to read a recently published paper to discover the latest development of deep learning.
However this beginner might have a problem in understanding the latest paper by jumping from the most basic theory to a much more sophisticated method, thus he/she would have to read more articles to help digest the latest paper.
Finding out the requisite papers could be tedious.
To help users achieve a comprehensive understanding of the domain,
PEG provides a graphic overview of the domain by depicting the relationships between the research branches and the development of each branch.

\emph{Search by single paper:}
In most cases, academic papers are not grown our of thin air. The majority of research works are done based on the previous studies. When reading a new paper, a beginner might want to find out how the idea of the paper was formed step by step from the very beginning, and whether there are any works that improve on the technique in the paper.
Our system is able to generate a PEG that explicitly shows the development of the query-paper, which will not only make it easier for users to find out former relevant papers,
but also lead users to the later papers which are closely related to the query-paper.

\emph{Search by two papers:}
Sometimes users are interested in discovering the relationship between two papers. For example, one might want to find out the relationship between a paper that proposed a classical theory and a recent paper that uses a variant of the theory to solve a specific problem.
Our system is able to present a PEG that shows a clear connection between the two papers, providing users with an idea of how the subject progressed step by step from the classical paper to the latest one.
PEG might also be useful for users who are curious about finding out the hidden connection between two papers that seem to be unrelated.

We believe the PEG can serve as an effective tool to help users navigate unfamiliar territory and discover previously unknown relationships between articles.
The main contributions of this paper are summarized as follows:
\begin{enumerate}
\item This paper proposes the concept of a paper evolution graph and formalizes the criteria for evaluating evolution graphs.
\item This paper supports three types of queries and provides efficient methods to construct evolution graphs given  different types of queries.
\item This paper integrates user preferences into the framework to generate graphs describing the multi-view relationships among articles.
\end{enumerate}

The remainder of this paper is organized as follows.
Section~\ref{related work} gives a survey of the related work.
Section~\ref{PEG construction overview} gives an overview of our approach for constructing a PEG.
Section~\ref{Identifying the topic distribution of papers} details the process of identifying the topic distribution of papers by metagraph factorization.
Section~\ref{Generating evolution graphs} describes the process of generating coherent evolution chains.
The experimental results and evaluation are reported in Section~\ref{Experiments} and Section~\ref{Evaluation}, respectively.
In Section~\ref{Conclusions and future work}, we conclude our work and give some ideas for future work.

\section{Related work}
\label{related work}
The problem of constructing a PEG relates to three aspects: document retrieval, organization of retrieval results, and topic discovery.
\subsection{Document retrieval}
The growing number of documents on the Web has accentuated the need for improving retrieval methods.
The probability ranking principle (PRP)~\cite{robertson1977probability} forms the bedrock of information retrieval.
It aims to achieve optimum retrieval by estimating the probability of relevance for each document (with respect to the current query) and ranking the documents according to the decreasing values of the probability of relevance.
There is also a rise of use of language models in information retrieval~\cite{lafferty2001document,lavrenko2001relevance}.
In the language modeling approach, each document is viewed as a language sample and a query is treated as a generation process.
The retrieved documents are ranked according to the probabilities of
generating a query from the corresponding language models of
these documents.
When the query has multiple interpretations, or there are multiple subtopics,
systems are expected to balance relevance and diversity~\cite{chen2006less,agrawal2009diversifying}.
The basic premise of result diversification is that the relevance of a set of documents depend not only on the individual relevance of its members, but also on how they relate to each another.
Maximizing diversity is especially useful in the feedback-relevant retrieval systems and commercial websites~\cite{shen2005active,yu2014latent}.



When it comes to academic literature, a paper's citation count is widely used in evaluating the importance of a paper since it has been shown
to strongly correlate with academic literature impact~\cite{narin1976evaluative}.
The Thomson Scientific ISI Impact Factor (ISI IF) is a representative approach using a paper's citation count,
and is defined as the mean number of citations to articles published in a journal over a two-year period~\cite{garfield1979citation}.
However, citation counting has well known limitations: citing papers with high impact and ones with low impact are treated equally in standard citation counting.
Google's PageRank algorithm counts not only the number of hyperlinks to a page.
It also computes the status of a Web page based on a combination of the number of hyper links that point to the page and the status of the pages that the hyperlinks originate from~\cite{brin1998anatomy}.
Papers with more citations are generally ranked higher, and they get a further boost if they are referenced by highly cited articles~\cite{butler2004science}.~\cite{chen2007finding} applied the PageRank algorithm to the scientific citation networks. They found out that according to PageRank model, some classical articles in the physics domain have a small number of citations but also have a very high PageRank.


\subsection{Retrieval result organization}
Search engines and recommendation systems play a crucial role in paper retrieval.
However, most of them are limited to list-output models, i.e., the retrieval results are listed one by one and isolated from each other.
Although these systems display useful information, simply listing the output is not sufficient for users to capture the relationships among retrieval results.
There are a few systems that move beyond the list-output model.
For example, Web of Science creates a citation map for each query article based on its forward and backward citation relationship.
In the topic detection task,~\cite{jo2011web} aimed to discover the evolution of topics over time in a paper collection.
The discovered topics were connected to form a topic evolution graph using a measure derived from the underlying paper network.
Graph-based and network-based models are also used to represent and analyze the relationships
among scientific authors. For example,~\cite{newman2001scientific} and ~\cite{tang2008arnetminer} utilized the
publications of authors to analyze and visualize co-author and citation relationships in the scientific literature.

Representing the retrieval results in a structured way has attracted more attention beyond the academic literature retrieval domain.
The ostensive browsing model~\cite{campbell2000interactive} use paths and nodes to represent interactive feedback-relevant searching process, where users move from node (information object) to node via links(accessibility relationships).
The path is a sequence of nodes for users to trace and explore.
In the news analysis domain, numerous works have proposed different notions of storylines~\cite{ahmed2011unified,yan2011evolutionary,allan2001temporal,shahaf2010connecting}.
In addition, graph representations are common across a variety of related problems.
For example,~\cite{kleinberg2003bursty} focused on discovering bursty and hierarchical structures in text streams.~\cite{makkonen2003investigations} suggested modeling news topics in terms of their evolving events.~\cite{mei2005discovering} aimed to discover and summarize the evolutionary patterns of themes in a text stream.

In a work most related to ours,~\cite{nallapati2004event} proposed a process called event threading, in which the structure of news events and their dependencies in a news topic were captured to generate a graph structure.~\cite{shahaf2012trains} created an evolution map for news events. They proposed the notion of topic coherence for an evolution path.
Our work share the same advantages of the previous works in that we try to uncover the relationships between retrieval results. However, we focus on the academic domain, where more information such as author and citation can be utilized rather than just the content information.

\subsection{Topic discovery}
Due to its importance and great application potential, topic research in scientific literature has recently attracted rapidly growing interest~\cite{mei2006probabilistic,spiliopoulou2006monic,schult2006discovering}.
Many existing approaches for scientific literature topic detection model a paper as a bag of words~\cite{bolelli2009topic,e2009topic}. However, the bag-of-words model is only effective for discovering topics when papers share a large proportion of lexically equivalent terms. Several works integrated author information and content information to help detect topics~\cite{zhou2006topic,rosen2004author,steyvers2004probabilistic}.~\cite{he2009detecting} addressed the problem of topic detection by adapting the Latent Dirichlet Allocation model~\cite{blei2003latent} to the citation network.
In the work of~\cite{small1973co}, the relationship between two papers was measured via their co-citations.
In this work, we integrate the content, citation and author information to discover academic topics from different viewpoints.

\section{An overview of PEG construction }
\label{PEG construction overview}
This section presents the outline of the proposed PEG construction.
First, we give the definition of a paper evolution chain and a paper evolution graph.

$\textbf{Definition 1}$ (\textit{Paper evolution chain}). A paper evolution chain $\textit{L}$ of length $\textit{n}$ is a simple directed path with $\textit{n}$ vertices, denoted by $\textit{L}=(p_1,p_2\cdots p_n)$, where $\textit{p}_1\cdots \textit{p}_n$ is a sequence of chronologically ordered and topically cohesive papers.

$\textbf{Definition 2}$ (\textit{Paper evolution graph}). A paper evolution graph $G=(V, E)$ is a directed graph consisting of several evolution chains $\textit{L}_i$, denoted by $G=\mathbf{U}(\textit{L}_i)$, where each $\textit{L}_i$ focuses on different topics.

\begin{figure}[t]
  \centering
     {\includegraphics[width=1\linewidth]{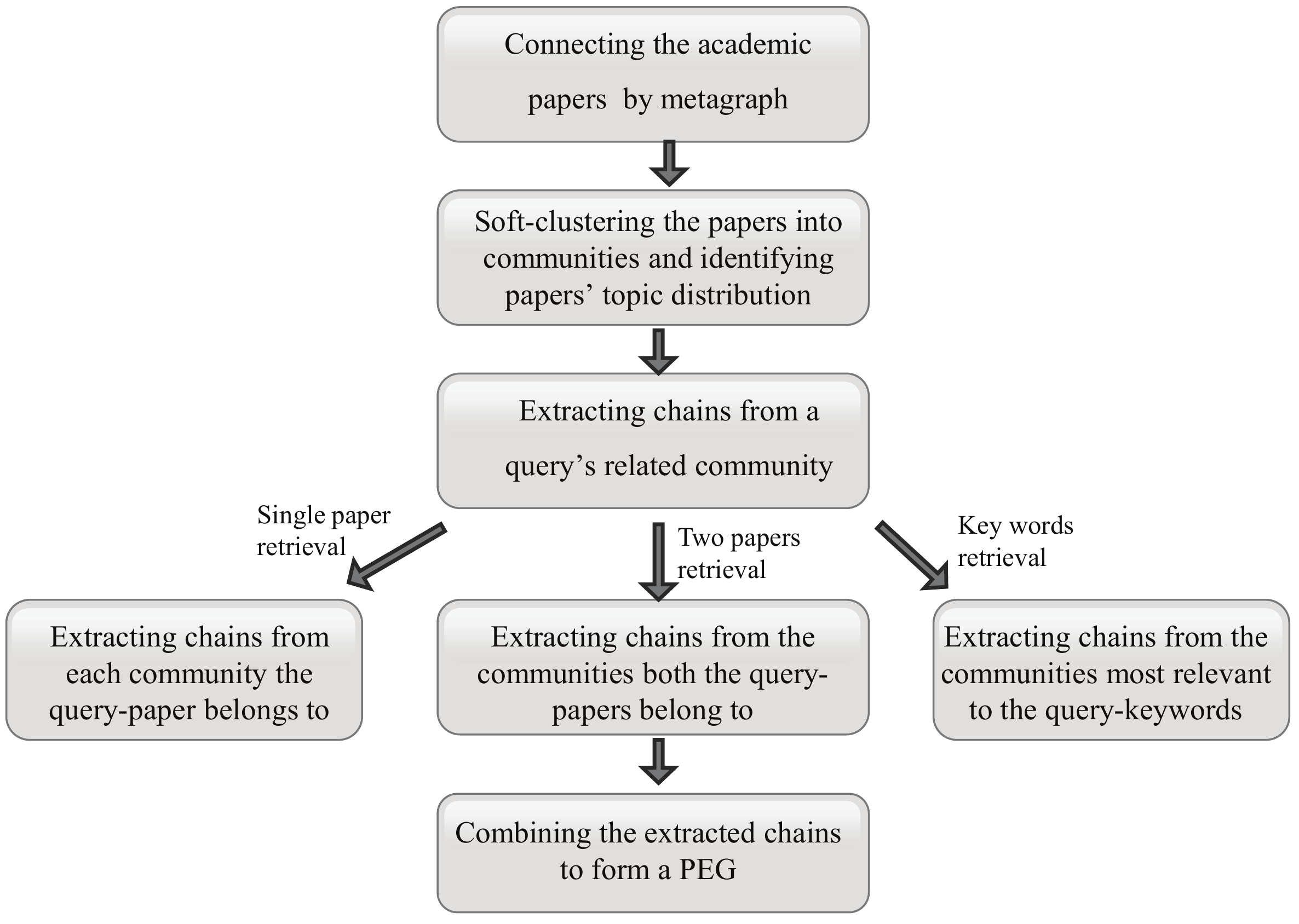}}
\caption{Framework of the PEG construction approach.}\label{fig:Framework of PEG construction approach}
\end{figure}

Fig.~\ref{fig:Framework of PEG construction approach} gives the procedure for constructing a PEG.
In the first step, we build the metagraph of our system specifying the relationships between ``word'', ``author'' and ``paper''.
A metagraph is a relational hypergraph representing multi-relational and multi-dimensional data~\cite{lin2009metafac}.
It is a graph with its nodes (called facets) representing the entities and its edges (called hyperedges) corresponding to the interactions between the nodes.
The metagraph of our system is shown in Fig.~\ref{fig:Metagraph of our system}.
A metagragh is different from the traditional graph in that each node/facet represents an ensemble of the entity.
For example, the author facet is a set of authors, and the paper facet is a set of papers.
The hyperedge connecting the facets represents the relationship between the entity sets.

We prepare three types of data according to three relationships in the metagraph, respectively:
the ``Content'' relationship between paper facet and word facet;
the ``Publish'' relationship between paper facet and author facet;
and , the ``Citation'' relationship between paper facet and paper facet.
Each relationship corresponds to an observed data.

The second step is to soft-cluster the papers in the dataset into communities according to the papers' topic distribution, which is achieved by multi-relational factorization of the metagraph.
In this step, each paper can be assigned to one or more communities.

The third step is to extract topically cohesive chains from the communities that are relevant to the query.
In this step, we first define the topic coherence of a given chain of papers.
Then the most coherent chains are extracted from each of the query-relevant community.

For different types of queries, the definition of  query-relevant community varies slightly.
For a single-paper query, the relevant communities are defined as the communities consisting of the query-paper.
For a two-paper query, the relevant communities are defined as the communities consisting both of the   query-papers.
For a keyword query, the query-relevant communities are the communities that include the papers relevant to the keyword.

After the most coherent evolution chains are extracted from the relevant communities, these chains are combined together to form a PEG in the last step.
Since each chain focuses on one aspect of the query, combining the chains gives us a comprehensive and holistic view about of evolution of the query.
\section{Identifying the topic distribution of papers}
\label{Identifying the topic distribution of papers}

To construct a PEG which fully covers the topics of a query, we should first identify the topic distribution of the papers in the dataset.
In this section, we introduce the approach to obtain the topic distribution of papers via metagraph factorization.
The first step is to build a metagraph that covers the papers' content, authorship and citation information.
Then the topic distribution of each paper is obtained by metagraph factorization.
\subsection{Constructing a metagraph}
In this work, each paper is modeled as a probabilistic mixture of topics, i.e., a paper belongs to one or more topics with different probabilities.
%
The topic distribution of a paper is defined as follows:

$\textbf{Definition 3}$ (\textit{Topic distribution}).  The topic distribution of a paper $\textit{p}$ is a nonnegative vector $T= (T_1,T_2,\ldots T_C)$,
where $C$ is the number of topics, $T_i$ is the probability that paper $\textit{p}$ belongs to the $\textit{i}$-th topic. $\sum_i T_i=1$.

\begin{figure}[t]
  \centering
    {\includegraphics[width=1\linewidth]{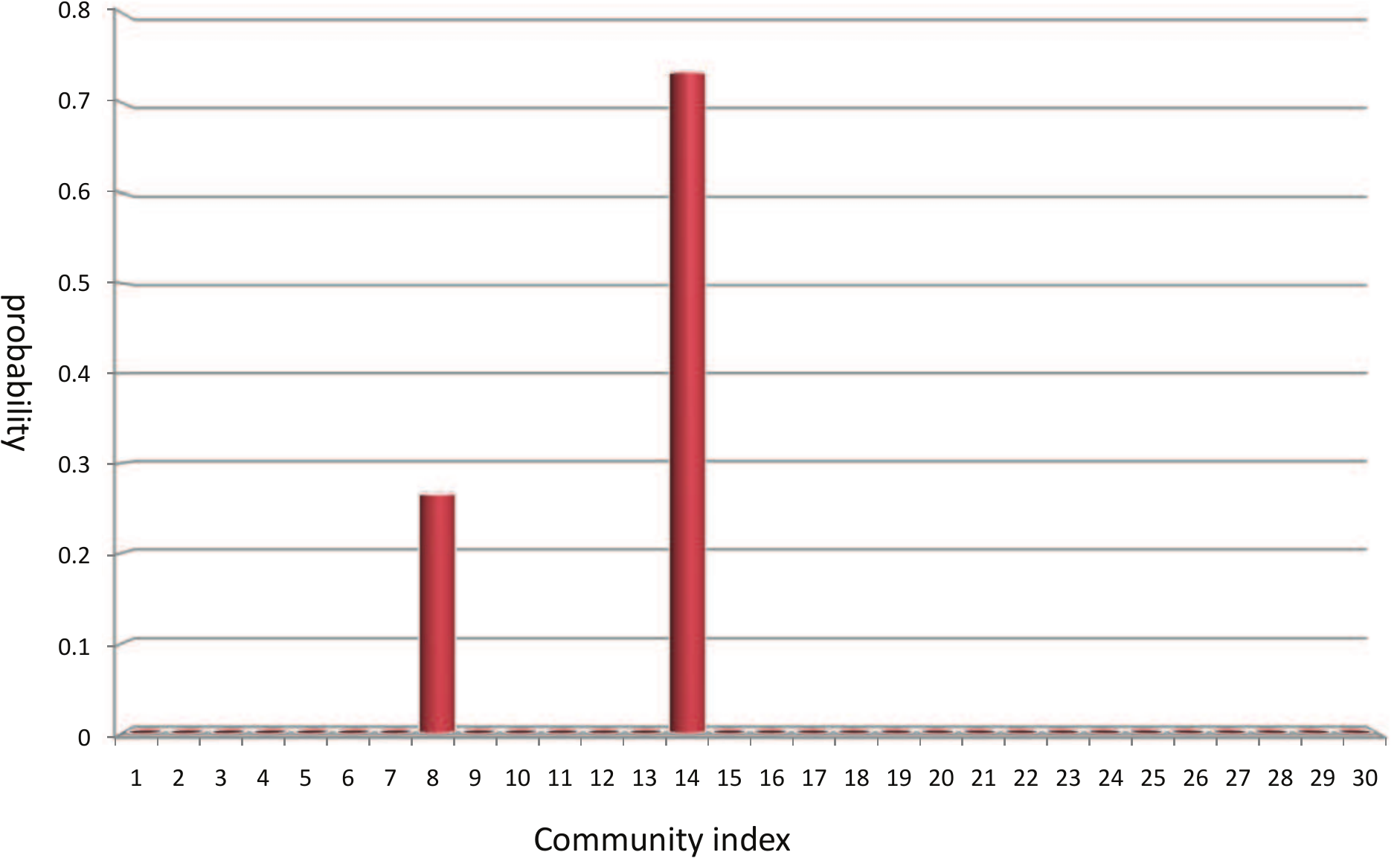}}
\caption{An example of topic distribution.}\label{fig:Topic}
\end{figure}

An example of topic distribution is shown in Fig.~\ref{fig:Topic}, where the x-axis denotes the index of topics and the y-axis is the probability of a paper belongs to a specific topic.
In this figure, $T_8=0.26$ and $T_{14}=0.74$.
In our approach, topic distribution is calculated by metagraph factorization-based clustering, in which similar articles  are grouped to the same topics.
Three types of papers are likely to share similar topics: papers that are similar in content; and papers that share the same authors; papers that have citation relationships.
To fully uncover the relationships among papers, we utilize three types of article information in the clustering step:
\begin{enumerate}
\item Content information.
The content of a paper conveys its topic in a most direct way.
Papers with high content similarity (e.g., word vector based similarity) will have similar topics.
\item Authorship information.
Since the research interests of a researcher are limited, papers published by the same author are likely to focus on the same topic.
\item Citation information.
If a paper cites another paper, there is a high probability that the two papers have the same topic.
\end{enumerate}
The relationships between papers can be measured in the paper-word space, paper-author space, and paper-paper space.
The three spaces are not independent. In fact, they share one common dimension: the paper dimension.
The relationships between the spaces can be represented by a metagraph, as shown in Fig.~\ref{fig:Metagraph of our system}.

Let $V$ and $R$ denotes the set of facets and edges respectively, where $v^{(i)}$ denotes the $\textit{i}$-th facet, and $e^{(r)}$ represents the $\textit{r}$-th edges.
A hyperedge/relation $e^{(r)}$ is said to be incident to a facet $v^{(q)}$ if $v^{(q)} \in e^{(r)}$.
There are three facets $V=\{v^{(i)}\}, i=1,2,3$ and three relationships $E=\{e^{(j)}\}, j=1,2,3$ in Fig.~\ref{fig:Metagraph of our system}.
The facets (paper, author and word) are connected by hyper-edges $e^{(1)}$, $e^{(2)}$ and $e^{(3)}$.
Thus {$e^{(1)}=\{v^{(1)},v^{(1)}\}$ represents the citation relationship between papers;
$e^{(2)}=\{v^{(1)},v^{(2)}\}$ represents the content relationship between papers and words;
and $e^{(3)}=\{v^{(1)},v^{(3)}\}$ represents the publishing relationship between authors and papers.

By constructing the metagraph, we are able to integrate the content, author and citation information, which were originally independent.
\subsection{Soft-clustering papers into communities}
\begin{figure}[t]
  \centering

     {\includegraphics[width=0.9\linewidth]{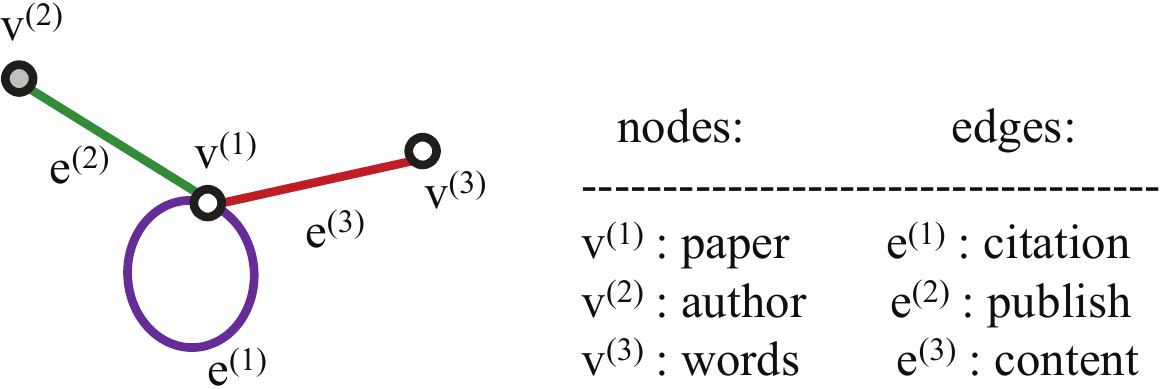}}

\caption{The metagraph of our system.}\label{fig:Metagraph of our system}
\end{figure}
A clustering technique is  crucial for large-scale topic discovery.
Since there are three relationships in our system, a  multi-relational clustering technique is required.
Clustering entities with multiple relationships considers joint factorization over two or more matrices.
There are numerous works addressing multi-relational clustering~\cite{zhu2007combining,long2006spectral,lin2009metafac, banerjee2007multi}.
In this paper, we apply a metagraph factorization method proposed by~\cite{lin2009metafac} to soft-cluster the papers.
 This approach is a fast and practical approach to extract communities from multiple-relationships on the basis of tensor operation.
 Furthermore, the approach is incremental in that it can be readily modified to deal with time evolving data, where the relational data is modeled as evolving tensor sequences, which makes the technique scalable.
\subsubsection{Background knowledge of the tensor}
This subsection provides the background knowledge on the tensor and the operations used in this work.
A tensor is a mathematical representation of a multi-way array.
The order of a tensor is the number of modes (or ways).
For example, a first-order tensor is a vector, a second-order tensor is a matrix, and a
third-order tensor is a cube.
In this work, we use $\textbf{x}$ as a vector, $\textbf{X}$ as a matrix, and $\mathcal{X}$ as a tensor.
The dimensionality of a mode is the number of elements in that mode.
For example, a nonnegative tensor $\mathcal{X}\in \Re^{I_1\times I_2\times I_3}$ has three modes with dimensionalities of $I_1$, $I_2$ and $I_3$, respectively.
 Tensor factorization or multi-linear matrix factorization is widely used in recommender systems~\cite{yu2015multilinear}.
The basic tensor operations we used in this work includes:

\emph{Mode-d unfolding}:
Unfolding is a process of reordering the elements of an M-way array into a matrix.
The mode-d unfolding of a tensor $\mathcal{X}\in R^{I_1\times I_2\times \cdots\times I_N}$ is denoted by $\textbf{X}_{(d)}$, i.e., $unfold(\mathcal{X},d)=\textbf{X}_{(d)}\in R^{I_d\times \prod_{q\in {1\cdots M},q\neq d}I_q}$. Unfolding a tensor on mode $d$ returns a matrix with $I_d$ rows. Its column number is  the product of dimensionalities of all the modes except mode $d$.
The inverse operation is denoted as $\mathcal{X}=fold(\textbf{X}_{(d)})\in R^{_1\times I_2\times \cdots\times I_N}$.
Unfolding can be defined on two or more modes. For example, mode $c,d$ unfolding of a tensor is defined by $unfold(\mathcal{X},(c,d))=\textbf{X}_{(c,d)}\in R^{Ic\times I_d\times \prod_{q\in {1\cdots M},q\neq c,d}I_q}$ where $\textbf{X}_{(c,d)}$ is a three way tensor (a cube).

\emph{Mode-d product}:
The mode-d product $\mathcal{Y}=\mathcal{X}\times _d \textbf{A}$ of a tensor
$\mathcal{X}\in R^{J_1\times J_2\times \cdots\times J_N}$ and a matrix $\textbf{A}\in R^{I_n\times J_n}$ is a tensor $\mathcal{Y}\in R^{J_1\times \cdots\times J_{n-1}\times I_n\times J_{n+1}\times\cdots\times J_N}$.
Elementwise, we have $\mathcal{Y}_{j_1,j_2,\cdots,j_{n-1},i_n,j_{n+1},\cdots,j_N}=\sum_{j_n=1}^{J_n}g_{j_1,j_2,\cdots,j_N}a_{i_n,j_n}$.

\emph{Tensor vectorization}: Vectorization is the process of linearizing the elements of an M-mode array  into a vector and is denoted by $\textbf{x}=vec(\mathcal{X})$.

\emph{Khatri-Rao Product}: For two matrices $\textbf{A}=[\textbf{a}_1,\textbf{a}_2,\cdots, \textbf{a}_J]\in R^{I\times J}$ and
$\textbf{B}=[\textbf{b}_1,\textbf{b}_2,\cdots \textbf{b}_J]\in R^{T\times J}$ with the same numbers of columns $J$, their Khatri-Rao product, denoted by $\odot$, is defined by\\
$\textbf{A}\odot \textbf{B}=[vec(\textbf{b}_1\textbf{a}_1^T)  \ vec(\textbf{b}_2\textbf{a}_2^T)\cdots \ vec(\textbf{b}_J\textbf{a}_J^T)]\in R^{IT\times J}$.

\emph{Mode-d accumulation}: A mode-d accumulation of a tensor $\mathcal{X}$ is defined as $acc(\mathcal{X},d)=\textbf{X}_{(d)}\textbf{1}\in R^{I_d }$.

Accumulating a tensor on mode $d$ can be calculated by unfolding the tensor on mode $d$ into a matrix and then multiplying the matrix with an all-one vector (summing up all the columns).
Accumulation on two modes $c$ and $d$ is defined by $acc(\mathcal{X},(c,d))=\textbf{X}_{(c,d)}\times_3\textbf{1}\in R^{I_c\times I_d}$.
Readers can refer to~\cite{bader2006algorithm} for a more comprehensive review of tensors.
\subsubsection{Clustering the papers into communities}
Since the hyperedges in our system are all two-way relationships,
a set of observations is represented as a two-way tensor, i.e., a matrix.
Each relation $e^{(j)}$ forms a matrix and corresponds to an observed data.
The``Citation'' hyperedge corresponds to a second-order tensor $\mathcal{X}^{(1)}\in R^{P\times P}$ where $P$ is the number of papers.
The``Content'' hyperedge corresponds to a second-order tensor $\mathcal{X}^{(2)}\in R^{P \times W}$, where $W$ is the number of words.
The``Author'' hyperedge corresponds to a second-order tensor $\mathcal{X}^{(3)}\in R^{P \times N}$ where $N$ is the number of authors.

The relationship between any two members $\textit{i}$ and $\textit{j}$ in a community $\textit{k}$ is denoted as $x_{ij}$.
Let $p_{k \rightarrow i}$ indicates how likely an interaction in the $\textit{k}$-th community involves the $\textit{i}$-th member and $\textit{p}_k$ indicates the probability of an interaction in the $\textit{k}$-th community.
$x_{ij}$ can be represented by $x_{ij}\approx \sum_k p_k \cdot p_{k\rightarrow i}\cdot p_{k\rightarrow j}$.
A set of such relationships among entities in facet $v^{(a)}$, $v^{(b)}$ can be written by:
\begin{equation}
\label{tensor formulate}
\mathcal{X}\approx \sum _{k=1}^K p_k\circ u_k^{(a)}\circ u_k^{(b)}= \mathbf{P} \times_1 \mathbf{U}^{(a)}\times_2 \mathbf{U}^{(b)}.
\end{equation}

The data tensor $\mathcal{X} \in \Re _+ ^{I_a \times I_b}$ represents the observed two-way interactions among the facets $v^{(a)}$, $v^{(b)}$ and $K$ is the number of communities.
$\mathbf{U}^{(q)}$ is an $I_q \times K$ matrix, where $I_q$ is the size of $v^{(q)}$.
$p_{k\rightarrow i_q}$ is the $(i_q,k)$-element of $\mathbf{U}^{(q)}$ for $q=a,b$.
$\mathbf{P}$ is a diagonal matrix with the diagonal elements representing the probabilities of each community, i.e., $p_k=\mathbf{P}(k,k)$.

Equation~\ref{tensor formulate} can be viewed as community discovery in a single relation.
Since there are three relations in our system,
our objective is to factorize all the data tensors such that all tensors can be approximated by a common nonnegative core tensor $\mathbf{P}$ and a shared nonnegative factor $\textbf{U}^{(1)}$, i.e. to minimize the following cost function:
\begin{equation}
\label{target}
\begin{split}
J(G)=\mathop{\min}\limits_{z, U^{(q)}} w_1 \times D(\mathcal{X}^{(1)}\| \mathbf{P} \times_1 \mathbf{U}^{(1)}\times_2 \mathbf{U}^{(1)}) \\
+ w_2 \times D(\mathcal{X}^{(2)}\| \mathbf{P} \times_1 \mathbf{U}^{(1)}\times_2 \mathbf{U}^{(2)}) \\
+ w_3 \times D(\mathcal{X}^{(3)}\| \mathbf{P} \times_1 \mathbf{U}^{(1)}\times_2 \mathbf{U}^{(3)})\\
s.t. \mathbf{P} \in \Re_+^{K\times K}, \mathbf{U}^{(q)}\in \Re _+ ^{I_q \times K} \forall q, \sum_i \mathbf{U}_{ik}^{(q)}=1 \ \forall q\forall k.
\end{split}
\end{equation}
Where $K$ is the number of communities. $D(\cdot\|\cdot)$ is the KL-divergence and $w_r$ is the weight of $\mathcal{X}^{(r)}$.


We apply the tensor operation-based metagraph factorization algorithm developed in~\cite{lin2009metafac} to find a local minima solution.
The solution shares the same form of the expectation-maximization algorithm and can be found by the following multiplicative updating algorithm:

In the first step, for each $e^{(r)}$, compute a tensor $\mathcal{C}^{(r)} \in \Re _+ ^{I_1^r\times I_2^r \times K}$ by
\begin{equation}
\label{target1}
\mu^{(r)}\leftarrow \textit{vec} (\mathcal{X}^{(r)} \oslash ([\mathcal{C}] \prod _{n:v^{(n)} \in e^{(r)}} \times_n \mathbf{U}^{(n)}))
\end{equation}
\begin{equation}
\label{target2}
\mathcal{C}^{(r)}=\textit{fold} (\mu^{(r)}\odot (\mathbf{c}\odot \mathbf{U}^{(2)}_{(r)} \ast \mathbf{U}^{(1)}_{(r)})^T)
\end{equation}
Where $\oslash$ is the element-wise division.

In the second step, $\mathbf{P}$ and $\mathbf{U}^{(q)}$ are updated by:
\begin{equation}
\label{target3}
\mathbf{P} \leftarrow \frac{1}{3} \sum_{r\in E} \textit{acc} (\mathcal{C}^{(r)},3)
\end{equation}
\begin{equation}
\label{target4}
\mathbf{U}^{(q)}\leftarrow \sum_ {l:e^{(l)}\in v^{(q)}} \textit{acc} (\mathcal{C}^{(l)},(3,q))
\end{equation}

Equation~\ref{target1} and Equation~\ref{target2} correspond to the E-step and Equation~\ref{target3} and Equation~\ref{target4} correspond to the M-step.
The information in each data tensor is aggregated at the E-step and is shared by the core tensor and all facet factors at the M-step.
Algorithm~\ref{table1} summarizes the process of metagraph factorization.

After multi-relational factorization, the topic distribution $T=(p(1|i),\cdots p(k|i),\\ \ldots p(K|i))$ of the $\textit{i}$-th paper in the dataset is calculated by $p(k|i)=p(i|k)\times p(k)/p(i)$, where $p(i|k)$ is the $(i,k)$-element of $U^{(1)}$ which denotes how likely it is that the $\textit{k}$-th community includes the $\textit{i}$-th paper.
$p(i)$ is the probability of a relation involving paper $i$ and is defined as $p(i)=\sum _{k}p(i|k)p(k)$.

After the topic distribution is acquired, each paper is assigned to several (usually $1$ to $3$) communities, which describe different topics of the paper.
The $\textit{i}$-th paper in the dataset is considered to belong to a community $k$ if $p(k|i)\geq Com_t$ where $Com_t$ is a threshold parameter ($Com_t=0.2$ in this paper).
For example, the paper with a topic distribution shown in Fig.~\ref{fig:Topic} belongs to community $8$ and community $14$.
Users can tune down $Com_t$ to retrieve papers from more communities when a PEG with more diversity is desired.
\begin{algorithm}[h]
\caption{Metagraph Factorization}
\label{table1}
\begin{algorithmic}
\REQUIRE ~~\\                          

    metagraph $G =(V,E)$ and data tensors $\mathcal{X}^{(1)}$, $\mathcal{X}^{(2)}$,$\mathcal{X}^{(3)}$  on $G$;\\

\ENSURE ~~\\                           

    $\textbf{c}$ and $\mathbf{U}^{(1)}$, $\mathbf{U}^{(2)}$, $\mathbf{U}^{(3)}$;\\
\JUAN ~~\\
      Initialize $c$ and $\mathbf{U}^{(1)}$, $\mathbf{U}^{(2)}$, $\mathbf{U}^{(3)}$;\\
                  Repeat until \textit{convergence};\\
                  \FOR {each $r$ $\in$ $E$}
                    \STATE Compute $\mathcal{C}^{(r)}$ by equation~\ref{target1} and equation~\ref{target2};\\
                   \ENDFOR

            \STATE Update $c$ by equation~\ref{target3};\\
            \FOR {each $q\in V$}
                    \STATE update $\mathbf{U}^{(1)}$, $\mathbf{U}^{(2)}$, $\mathbf{U}^{(3)}$ by equation~\ref{target4};\\
                   \ENDFOR

\end{algorithmic}
\end{algorithm}
\section{Generating evolution graphs}
\label{Generating evolution graphs}
In this section, we describe our approach for extracting topically cohesive chains and constructing PEGs.
In order to extract cohesive chains, we first define the link strength between two papers, based on which we define the topic coherence of a chain.
Then we extract chains with the most coherent topic from the query-related communities.
After topically cohesive chains are extracted, they are combined to construct a PEG.

\subsection{Computing link strength between adjacent papers}
\label{calculating word influence between adjacent papers}
Given a chain of papers $L=(p_1 \cdots p_n)$, the link strength between adjacent papers can be measured by the similarity between the two papers.
Traditionally, papers are represented by vectors of term frequencies, i.e., using the ``bag-of-words'' model.
Each term is then assigned a ``weight of importance'' using a weighting metric such as the TF-IDF weighting scheme~\cite{salton1971smart}.
A simple measurement of the link strength is the word vector based similarity between papers,
e.g., the cosine distance between word vectors.
However, this type of similarity is not always informative for measuring the link strength.
First, since the bag-of-words only consists of terms that appear in a paper's original source text,
there exists two linguistic phenomena: ambiguity and synonymy~\cite{aljaber2010document}.
Ambiguity occurs when papers share lexically similar, but semantically distinct terms.
It can make papers appear more similar than they actually are.
Synonymy, on the other hand, occurs when two papers share semantically related, but lexically dissimilar words.
As a result, two correlative papers appear less correlative than they actually are.
Second, word vector-based similarity always considers each term to be of equal importance.
However, when calculating similarity under a certain topic, some terms are more significant than other terms.
For example, say there is a paper focuses on the NMF and its application to ``image compression''.
Another paper also applies NMF method, but uses it to address a data mining problem.
The two papers are relevant because they both use NMF to solve problems.
However, when the topic of a chain is image compression, these two papers should not be considered relevant, which means we should give a small weight to the term ``NMF'' when calculating paper similarity under the image compression topic.
In this paper, we take into consideration the word influence in similarity calculations,
i.e., the influence of a word $w$ in the relation of $p_i$ and $p_{i+1}$.
With word influence, the similarity between two papers can be obtained under different topics by assigning different weights to each word.

Among the several proposed methods for measuring word influence, the majority of them focus on directed weighted graphs (e.g., the web, social networks, citations), in which influence is considered to spread through the edges.
Methods such as PageRank~\cite{brin1998anatomy}, authority computation~\cite{kleinberg1999authoritative} and random graph simulations~\cite{kempe2003maximizing} all make use of the link structure.
In this paper, we utilize the influence calculating algorithm proposed by~\cite{shahaf2010connecting}, where word influence is obtained by random walk.
The algorithm overcomes the two drawbacks of word vector-based similarity.
It creates a network between all the papers and words, where the relation between two papers is acquired by the word influence propagating in the network.
This framework allows two papers to be linked through a word $\textit{w}$ even if the $w$ does not appear in the two articles.

To create the network, we first created a bipartite directed graph $B=(V,E)$.
The vertices $V=V_P \bigcup V_W$ correspond to papers and words.
Fig.~\ref{Bipartiate graph of ramdon walk} shows a simple bipartite graph: the squares on the left side represent four papers and the circles on the right side denote five words.
We added both edges $(w,p)$ and $(p,w)$ for each pair of word $\textit{w}$ and paper $\textit{p}$.
The weights of edges indicate the strength of the relevance between papers and words.
For each paper, the weights of the paper-to-word edges are assigned with their TF-IDF features.
Since the weights are considered to be random walk probabilities, they are normalized over all the words such that $\sum_iweight(p,w_i)=1$.
The word-to-paper weights are initialized in the same way as the paper-to-word edges, but normalized over the papers.

In order to calculate $influence(p_i,p_j|w_k)$, we first computed the stationary distribution for random walks starting from $\textit{p}_i$.
The probability $p_(p_j|p_i)$ can be calculated from $\textit{p}_i$ to $\textit{p}_j$ through the whole word set.
To specify the effect of $w_k$ on these walks,
a graph $B'$ which is the same with $B$ is constructed except that there is no way out of the node $w_k$, i.e. the weights of the word-to-paper edges of $w_k$ is set to 0.
Again, the stationary distribution on $B'$ is calculated starting from $p_i$.
This time the probability $p_(p_j|p_i )$ is computed without the influence of $w_k$.
We denote this probability as $p_{-w_k} (p_j|p_i )$.
The word influence $w_k$ between $p_i$ and $p_j$ is defined as the probability difference between the two distributions and can be calculated by
\begin{equation}
 influence(p_i, p_j|w)= p(p_j|p_i )-p_{-w_k} (p_j|p_i).
 \end{equation}

Fig.~\ref{Word influence} shows an example of the word influence between paper $\textit{d}$ and two other articles $\textit{d}_1$ and $\textit{d}_2$.

\textit{$d$: Constrained Nonnegative Matrix Factorization for Hyperspectral Unmixing}

\textit{$d_1$: Spectral and Spatial Complexity-Based Hyperspectral Unmixing}

\textit{$d_2$: Hyperspectral Unmixing via L1/2 Sparsity-Constrained Nonnegative Matrix Factorization}
\begin{figure}[t]
  \centering
      \subfigure[]{\label{Bipartiate graph of ramdon walk}
      \includegraphics[width=0.2\linewidth]{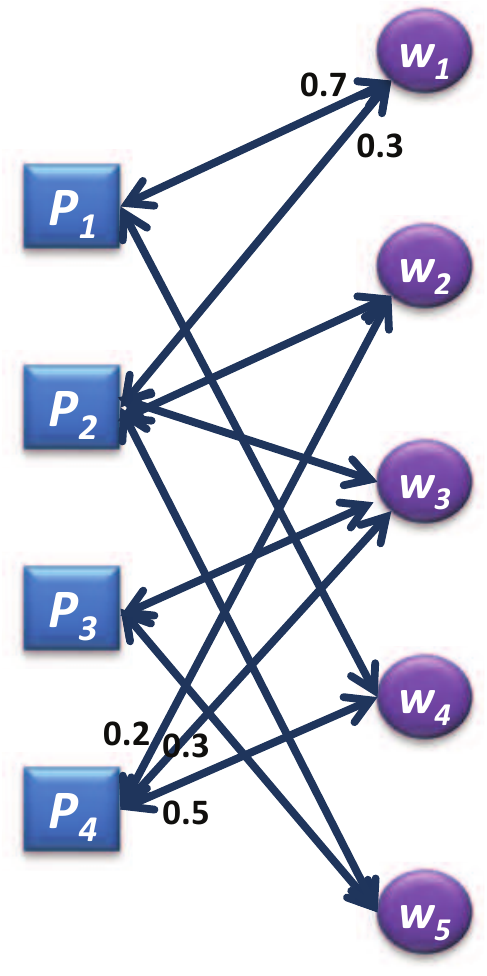}}
      \subfigure[]{\label{Word influence}
      \includegraphics[width=0.7\linewidth]{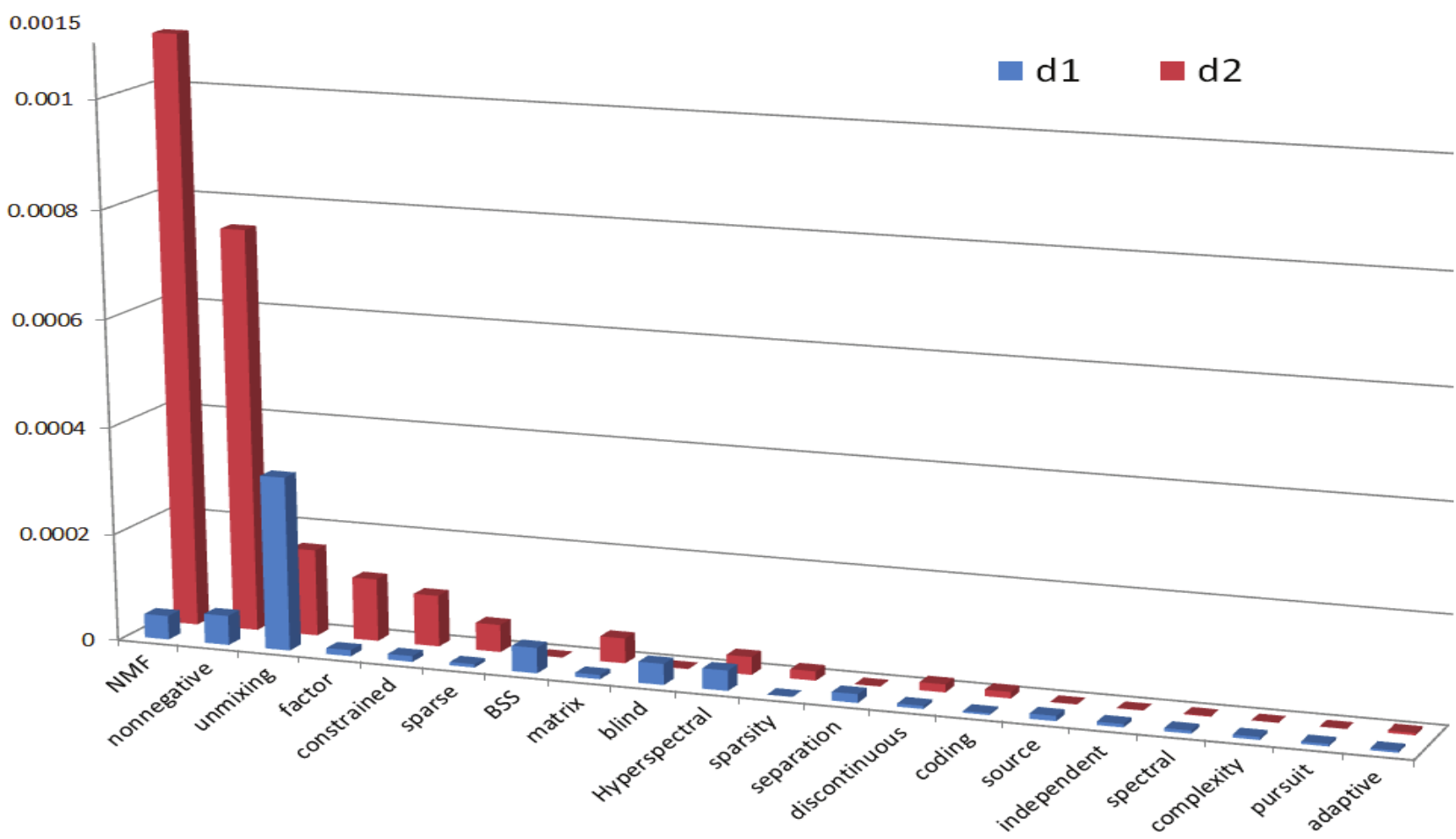}}
\caption{Word influence between papers.
(a) Word influence between $d$ and $d_1$, $d$ and $d_2$.
(b) Bipartiate graph of papers and words.}\label{fig:Word influence in paper transition}
\end{figure}
The horizontal axis corresponds to words and the vertical axis is the word influence.
The blue bars represent word influence for $d$ and $d_1$ while the red bars denotes the word influence for $d$ and $d_2$.
We can see that words such as ``unmixing,'' ``BSS,'' ``blind,'' and ``hyperspectral'' have higher influences in the relation between $d$ and $d_1$,
while ``NMF,'' ``nonnegative,'' and ``sparse'' have higher influence in the relation between $d$ and $d_2$.

After obtaining the word influence between each pair of papers in the dataset,
the similarity between two papers can be computed under different topics $T$ by assigning different weights to each word.
A topic $T_k$ is described by a vector $T_k=(t_1^k,\cdots t_n^k,\cdots t_W^k)$, where $t_n^k$ is the weight of word $w_n$ and $W$ is the number of words.
The similarity between two papers under a certain topic can be formulated as the sum of influences through all the words:
\begin{equation}
\label{similarity}
\begin{aligned}
 sim(p_i, p_j,T_k)=\sum _n t_n^k\times influence(p_i, p_j|w_n),\\
 ~s.t. \sum_n t_n^k=1.
 \end{aligned}
\end{equation}
\subsection{Evaluating coherence of chains}
\label{Extracting coherent chains using word influence}
In this paper, the coherence of the chain is measured by the strength of its weakest link due to the fact that a single poor transition can destroy the coherence of the entire chain~\cite{shahaf2010connecting}.
Given a chain of papers $L=(p_1,p_2, \cdots p_m)$, its coherence is defined as the maximum value of its weakest link strength among all the possible topics:
\begin{equation}
\label{coherenceInit}
\begin{aligned}
  cohere(L)=
  \mathop{\max}\limits_{T_k}\mathop{\min}\limits_{j=1,\cdots,m-1} sim(p_j,p_{j+1},T_k).
  \end{aligned}
 \end{equation}

To determine the topic coherence of a chain, we need to find a topic that maximizes the optimization problem in Equation~\ref{coherenceInit}.
However, Equation~\ref{coherenceInit} considers the topic of a chain to be constant.
In other words, the importance of the words is fixed while computing the link strength between adjacent papers.
In fact, the research topic is  always gradually changing over time.
A more reasonable objective function would consider the adjacent papers' similarity to be computed under a set of smoothly evolving topics $M=(T_1 \cdots T_{m-1})$.
In this case, the coherence of a chain is then defined as:
 \begin{equation}
 \label{finalobjective}
\begin{aligned}
  cohere(L)=&\mathop{\max}\limits_{M} \mathop{\min}\limits_{j=1,\cdots,m-1} sim(p_j,p_{j+1},T_j)\\
  &s.t. \sum_i t_i^j=1, \|t_i^j-t_i^{j+1}\| \leq r.
\end{aligned}
\end{equation}
where $t_i^j$ is the weight of the $\textit{i}$-th word in the topic $T_j$.  which is used to calculate the similarity between $p_j$ and $p_{j+1}$.
By introducing a range parameter $r$, the word importance (the topic) is allowed to change slightly between adjacent pairs of papers along the chain.
Our goal is to find a set of gradually changed topics such that the chain reaches its highest coherence.
Equation~\ref{finalobjective} can be readily formalized as a linear programming problem. Since the length of the chain and the number of total words are limited, the linear programming problem can be solved fast.

After selecting the most topically cohesive chains, these chains are combined to construct a PEG.
Since a paper can belong to $k (k>=1)$ communities after soft-clustering, chains extracted from different communities may share common nodes.
Fig.~\ref{fig:Combining chains to construct a graph} illustrates the process of combining two chains to form a graph.
When chains share the same nodes, the common nodes are merged.
The common nodes uncover the intersection between different technique routes.
\begin{figure}[t]
  \centering
     {\includegraphics[width=1\linewidth]{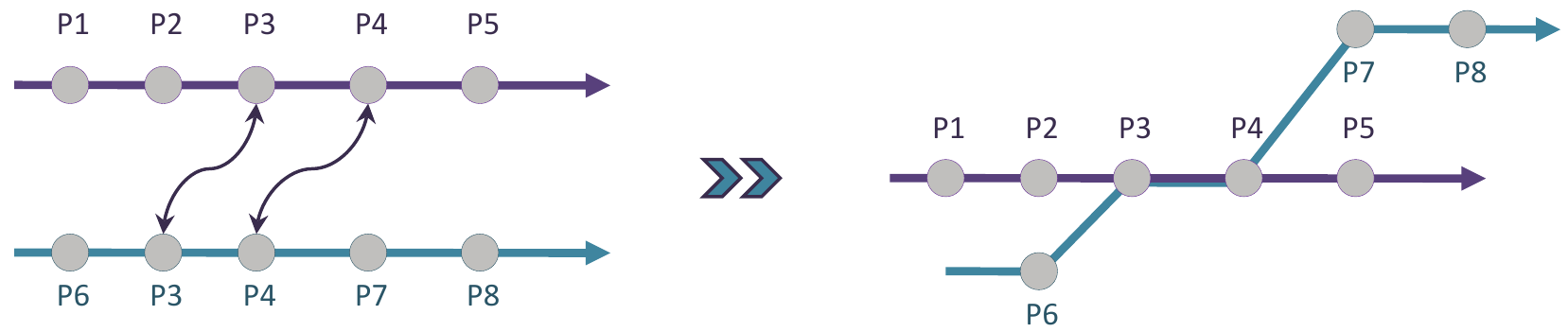}}
\caption{Combining chains to construct a graph.}\label{fig:Combining chains to construct a graph}
\end{figure}
\subsection{Chain extraction for three types of queries}
Our system supports three types of input queries: keyword, single-paper and two-papers queries.
For different types of queries, the PEG can be generated under the same framework with small variance.

Assume that the length of the evolution chain is set to $n$ either by our system or by the user.
When the query is a single paper $p$, our model generates a PEG discovering the evolution of $p$'s topics.
The first step is to find out the communities that include $p$ according to its topic distribution.
Then from each of $p$'s communities, our model selects a chain that has the strongest topic coherence from all the possible chains that contain $p$ as one of its nodes.
Finally, the chains selected from different communities are combined to form a PEG.

For two query-papers $p_s$ and $p_t$, our model constructs a PEG that uncovers the evolution relation between the papers.
Assume that $p_s$ was published before $p_t$.
The first step is to find out the shared communities of $p_s$ and $p_t$ according to their topic distribution.
Let $C_i (i=1\cdots R)$ denote the $\textit{i}$-th common community of $p_s$ and $p_t$, where $R$ is the number of shared communities.
For each $C_i$, our model compute the coherence of all the chains that start with $p_s$ and end with $p_t$ using Equation~\ref{finalobjective}.
Then a chain with the largest coherence is extracted from each $C_i$.
These chains are then combined together to construct a PEG.

When the query is a keyword, our system generates a PEG consisting of papers that are most relevant to the keyword by using the following steps:
\begin{enumerate}
\item Select $N$ papers that are most relevant to the query-keyword by the TF-IDF feature of the papers.
\item Divide the selected papers into groups according to the communities they belong to, i.e., papers that belong to the same community are assigned to the same group.
\item In each group, compute the topic coherence of all the possible chains of length $n$ and choose the most coherent chain.
\item Combine the chains to form a PEG.
\end{enumerate}
\subsection{Some issues on implementation}
\label{Some issues on implementation}

Two of the most time-consuming parts of the proposed model are: 1) hypergraph factorization to generate communities; 2) chain extraction from each query-related community.

The hypergraph factorization has a time complexity of O(n) per iteration.
The factorization can be readily modified to deal with time evolving data as proposed by~\cite{lin2009metafac}, where the relational data is modeled as evolving tensor sequences, which makes the technique well scalable.

Our method aims to search for the most coherent chains from each query-relevant community.
Since there are an enormous number of possible chains in a community, an exhaustive search method is not feasible when the dataset is very large.
To accelerate the computation, we narrow down the search region by selecting a small number of the most relevant papers as the candidates for constructing chains.

When the query is single paper $p$, we select $M$ papers, which are most similar with $p$ in each community that $p$ belongs to, using the similarity defined as:
 \begin{equation}
 \label{Closeness1}
      R(p,p_m)=\sum_k P(k)\times P(k\rightarrow p) \times P(k\rightarrow p_m)
\end{equation}
where $k$ is a community and the summation is taken over all the communities.

For two query-papers $p_s$ and $p_t$, $M$ papers that are most similar to the two papers are selected in each community which both $p_s$ and $p_t$ belong to, using the similarity defined as:
 \begin{equation}
 \label{Closeness2}
    \begin{aligned}
      R(p_s,p_t,p_m)&=\sum_k P(k)\times P(k\rightarrow p_m)  \\ &\times P(k\rightarrow p_s) \times P(k\rightarrow p_t).
      \end{aligned}
\end{equation}
In our experiment, $M$ is set to 50.

 Chain extraction involves candidate paper selection and the linear programming (Equation~\ref{finalobjective}) to determine the coherence of an evolution chain. The time complexity of candidate paper selection is O(n). Since the length of the chain and the number of total words are limited, the linear programming problem can be solved fast.

To discover whether the acceleration process will affect the final retrieval results, we conducted two sets of the experiments, where one set wass done with acceleration and the other without  acceleration.
We found out that in most cases the two groups of results are exactly the same, which means that the papers ruled out from the candidates do not contribute to the final results.

\section{Experiments}
\label{Experiments}
A real-world dataset of 24491 papers were collected to test the effectiveness of our structural retrieval approach.
Three types of queries were conducted on the dataset.
In addition, user preference was incorporated into our system to better meet users' needs.
\subsection{Dataset description}
We crawled and parsed the articles from the journal called \textit{IEEE Transactions on Geoscience and Remote Sensing} (TGRS) for the years 1980-2012 and a conference called \textit{IEEE International Geoscience and Remote Sensing Symposium} (IGARSS) for the years 1988-2012.
We chose this dataset because remote sensing is one of the research domain of our laboratory, which allows us to better analyze the retrieval results.
Each of the three relations (``Content,'' ``Author,'' and ``Citation'') in the dataset corresponds to a second-order tensor (a matrix).
The relations are summarized in Table~\ref{datasets}.
\begin{table}[h]
  \centering
  \begin{tabular}{ccc}
    \hline
    Relation & Tensor& Size\\
    ~&(incident facets)&~\\
     \hline
    Content & paper, word & $24491\times 27730$ \\
    Author & paper, author & $24491\times 38094$\\
    Citation & paper, paper & $24491\times 24491$ \\
    \hline
  \end{tabular}
  \caption{Summary of the relations in the TGRS dataset}
  \label{datasets}
\end{table}
The ``Content'' data measures the relationships between papers and words, where the words are extracted from the title, keywords, and abstract of each paper with stop words removed and stemming.
The TF-IDF (term frequency-inverse document frequency) metric is used to measure how important a word is to a paper.
For a paper $p$, its TF-IDF feature $\delta^p=(\delta_1^p, \delta_2^p,\cdots ,\delta_N^p)$ is calculated by
 \begin{equation}
  \delta_i^p=tf_{w_i,p} \times \log \frac{|P|}{\{ p' \in P | w_i \in p'\}}
  \end{equation}
where $|P|$ is the number of papers in the dataset and $tf_{w_i,p}$ is the frequency of the word $w_i$ in $p$.
In this paper, we use the number of times that word $w_i$ occurs in document $p$ as the word frequency.
The ``Author'' data is a 0-1 matrix with 1 referring to an author publishing a paper.
The ``Citation'' data is also a 0-1 matrix where 1 refers to a paper citing an other paper.

At the metagraph factorization stage, papers in the dataset are clustered into 30 communities.
Four of the communities are displayed as word clouds in Fig.~\ref{fig:wordclouds}.
The three relations are assigned with equal weights ($w_1=w_2=w_3=1/3$) in Equation~\ref{target} by default.
Users can also choose different weights according to their needs.

\subsection{Search by single paper}
For a single-paper query, we selected $50$ articles that were most similar to the query-paper in each of the query-paper's communities.
The similarity is measured by Equation~\ref{Closeness1}.
The selected papers are the candidates for constructing evolution chains.
In order to  analyze the retrieval results more accurately, we chose two subfields which our lab were familiar with to do the experiments: hyperspectral imagery classification and hyperspectral imagery unmixing.
Fig.~\ref{fig:PEG Single Paper classification} shows the PEG of the query-paper ``Spectral and Spatial Classification of Hyperspectral Data Using SVMs and Morphological Profiles.''
This article is about the application of morphological profiles in hyperspectral imagery classification using SVM as the classifier.
It belongs to community 11 and community 14 with probabilities of $0.65$ and $0.34$ respectively, thus the PEG is a combination of two chains of length 6.
As the word clouds shows, community 11 contains the keyword ``morphological,'' while community 14
includes articles relates to machine learning.
Correspondingly, the first chain of the PEG in Fig.~\ref{fig:PEG Single Paper classification} focuses on classification of hyperspectral imagery with the features relating to morphological profiles.
The second chain describes various support vector machine-based
classification methods.

\begin{figure}[t]
  \centering
      \subfigure[Word cloud of community 9.]{\label{fig:word2}\includegraphics[width=0.4\linewidth]{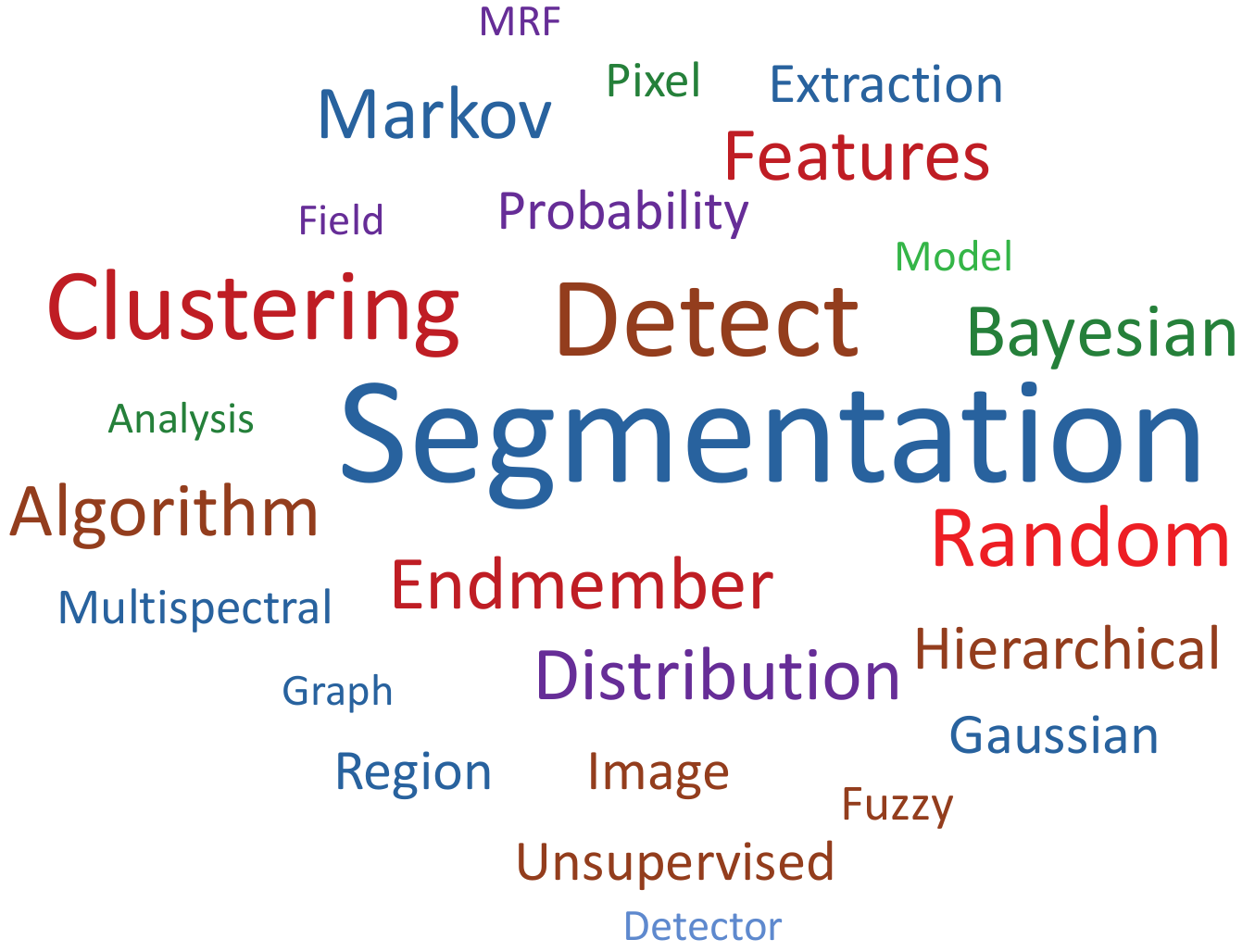}}
      \subfigure[Word cloud of community 11.]{\label{fig:word8}\includegraphics[width=0.4\linewidth]{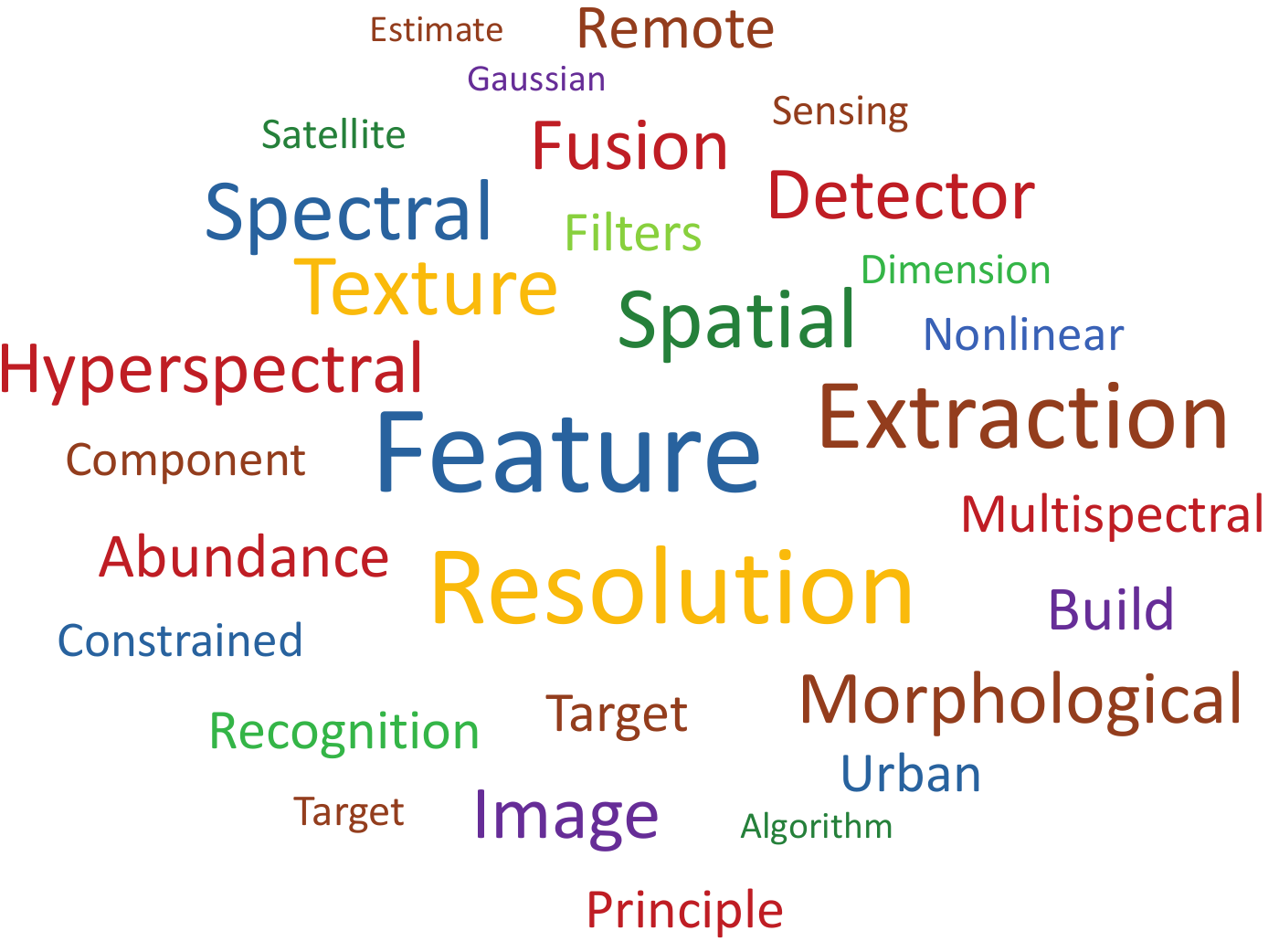}}
      \\
  \centering
      \subfigure[Word cloud of community 14.]{\label{fig:word11}\includegraphics[width=0.45\linewidth]{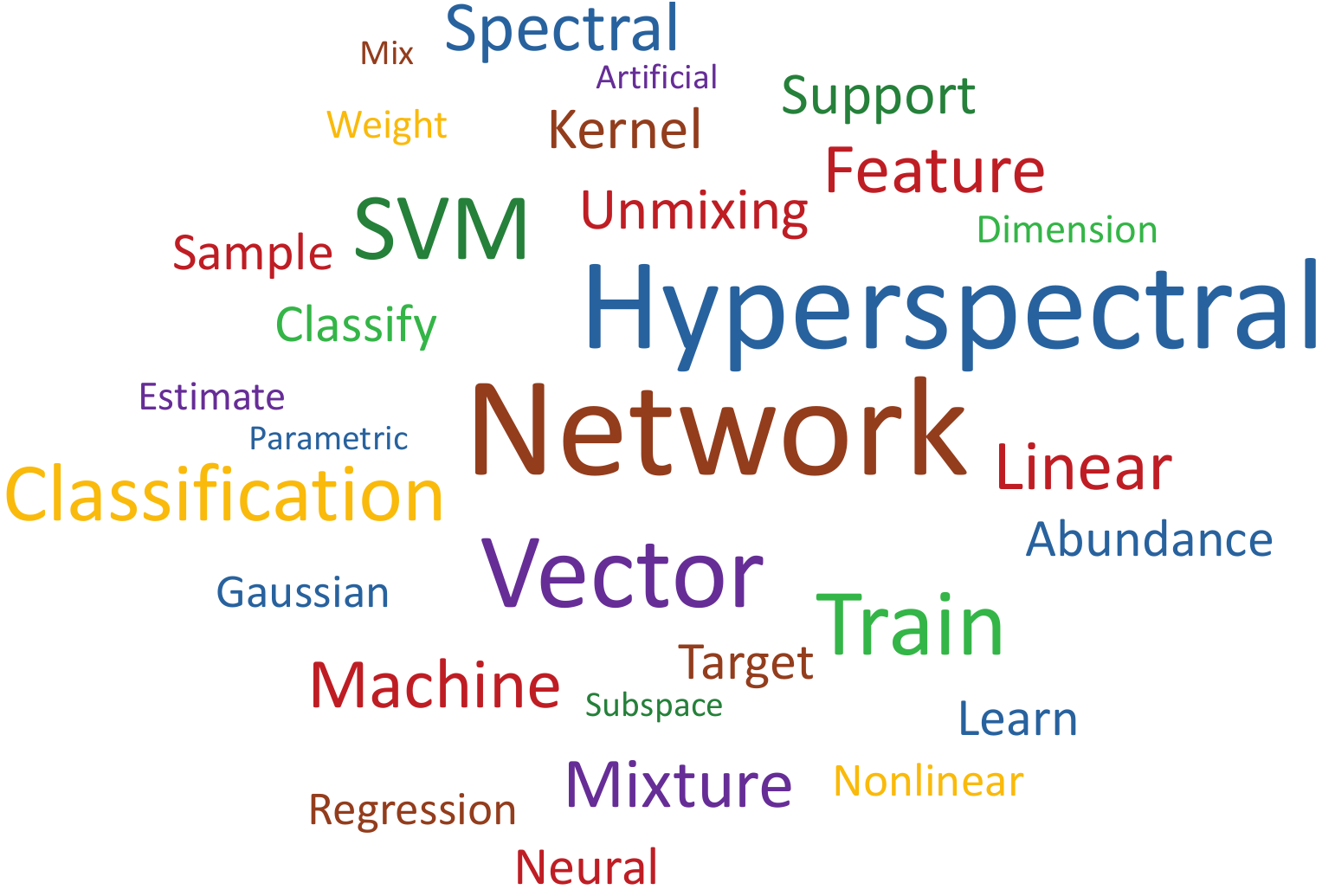}}
      \subfigure[Word cloud of community 30.]{\label{fig:word14}\includegraphics[width=0.4\linewidth]{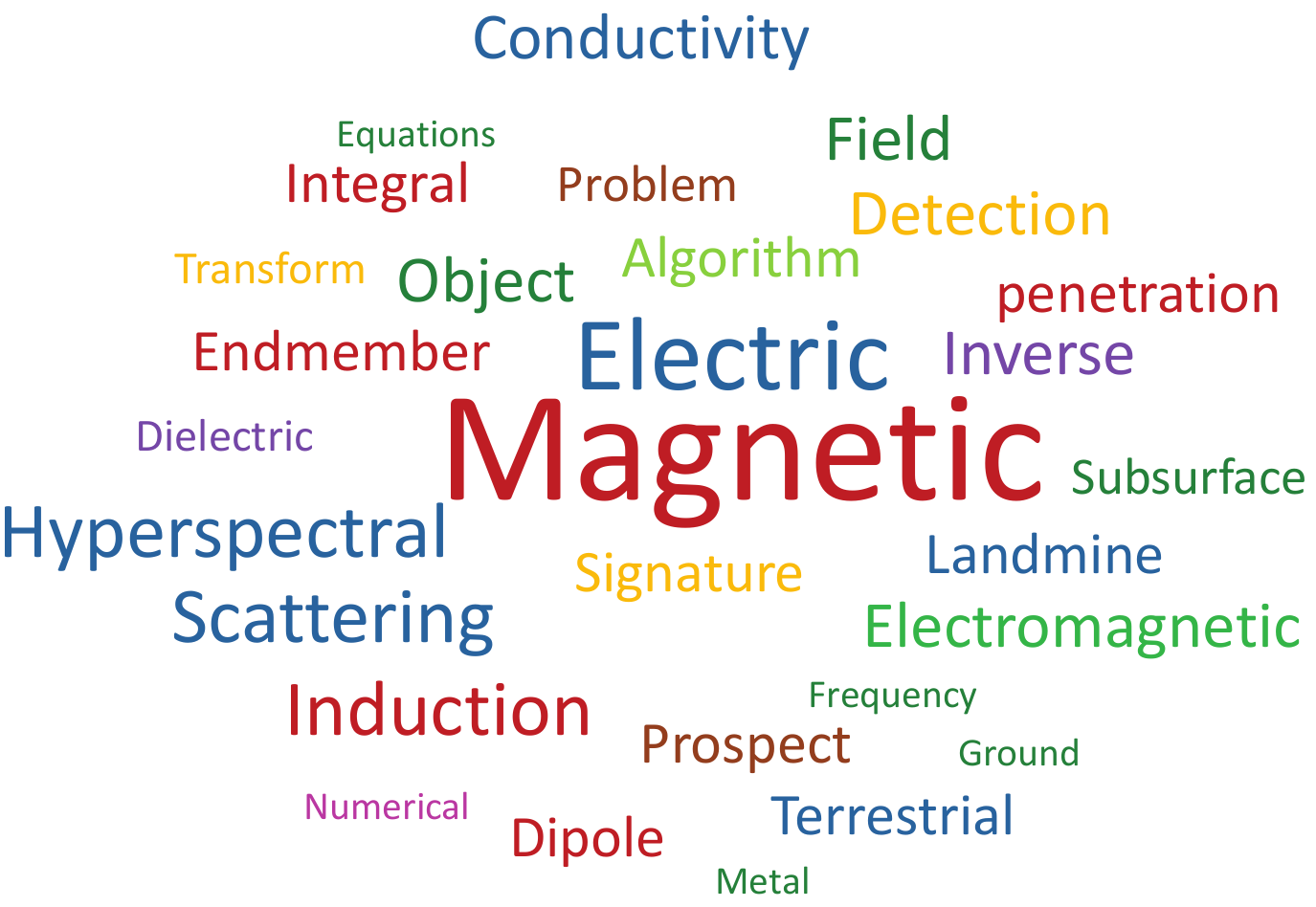}}
\caption{Word clouds for various communities}\label{fig:wordclouds}
\end{figure}

Fig.~\ref{fig:PEG Single Paper unmixing} displays the PEG of the query-paper ``Constrained Nonnegative Matrix Factorization for Hyperspectral Unmixing.''
The paper belongs to community 9 and community 14 with a probability of $0.36$ and $0.35$, respectively.
Thus, the PEG consists of two evolution chains. The length of the chain is set to $5$.
Chain 1 focuses on using nonnegative matrix factorization to unmix hyperspectral imagery with different constraints.
Chain 2 has two common papers with Chain 1, but it focuses more on  applying linear methods to do mixture analysis, which is a task substantially equivalent to the unmixing problem.
\begin{figure}[t]
  \centering

    {\includegraphics[width=1\linewidth]{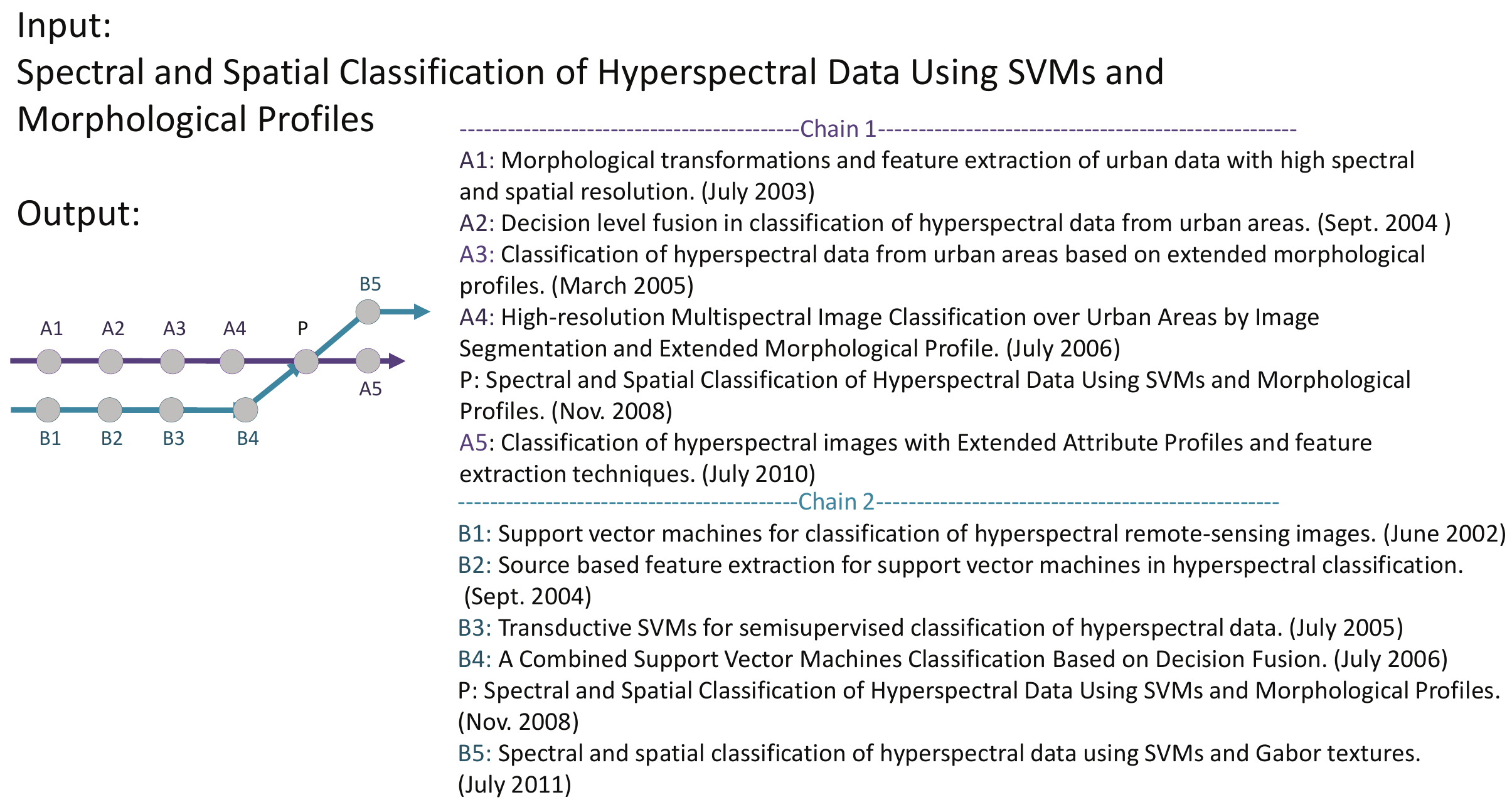}}

\caption{PEG of the query-paper ``Spectral and Spatial Classification of Hyperspectral Data Using SVMs and Morphological Profiles.''}\label{fig:PEG Single Paper classification}
\end{figure}
\begin{figure}[t]
  \centering

    {\includegraphics[width=1\linewidth]{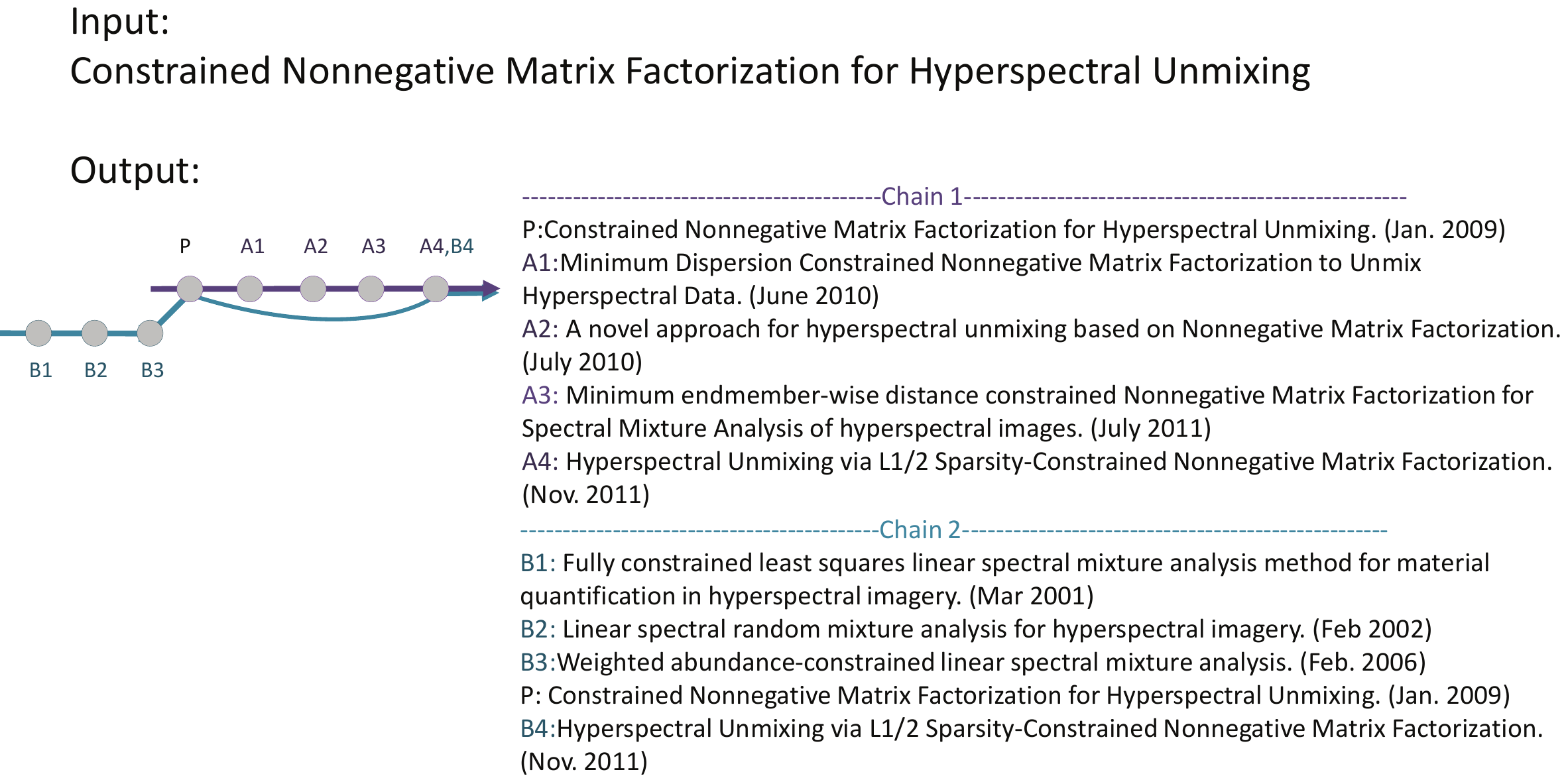}}

\caption{PEG of the query-paper ``Constrained Nonnegative Matrix Factorization for Hyperspectral Unmixing.''}\label{fig:PEG Single Paper unmixing}
\end{figure}
\subsection{Search by two papers}
For  two-papers retrieval, our system selected $50$ papers that were most relevant to the query-papers using Equation~\ref{Closeness1} in each of the communities shared by the query-papers.
%
Fig.~\ref{fig:PEG of two papers retrieval classification} shows the PEG
constructed for the query-papers ``Kernel-based methods for hyperspectral image classification'' and
``Kernel Nonparametric Weighted Feature Extraction for Hyperspectral Image Classification.''
The two papers both focus on hyperspectral imagery classification using kernel methods.
All papers in the chain apply kernel methods in hyperspectral imagery classification.

Fig.~\ref{fig:PEG of two papers retrieval unmixing} shows the PEG constructed for the query-papers
``Endmember Extraction From Highly Mixed Data Using Minimum Volume Constrained Nonnegative Matrix Factorization''
and ``A Novel Strategy of Nonnegative-Matrix-Factorization-Based Polarimetric Ship Detection''.
Both of the papers address hyperspectral imagery detection using nonnegative matrix factorization.
The former paper uses NMF to address a feature extraction task, which is a general detection problem,
while the latter paper extends the method to solve a specific ship detection problem.
\begin{figure}[t]
  \centering

     {\includegraphics[width=1\linewidth]{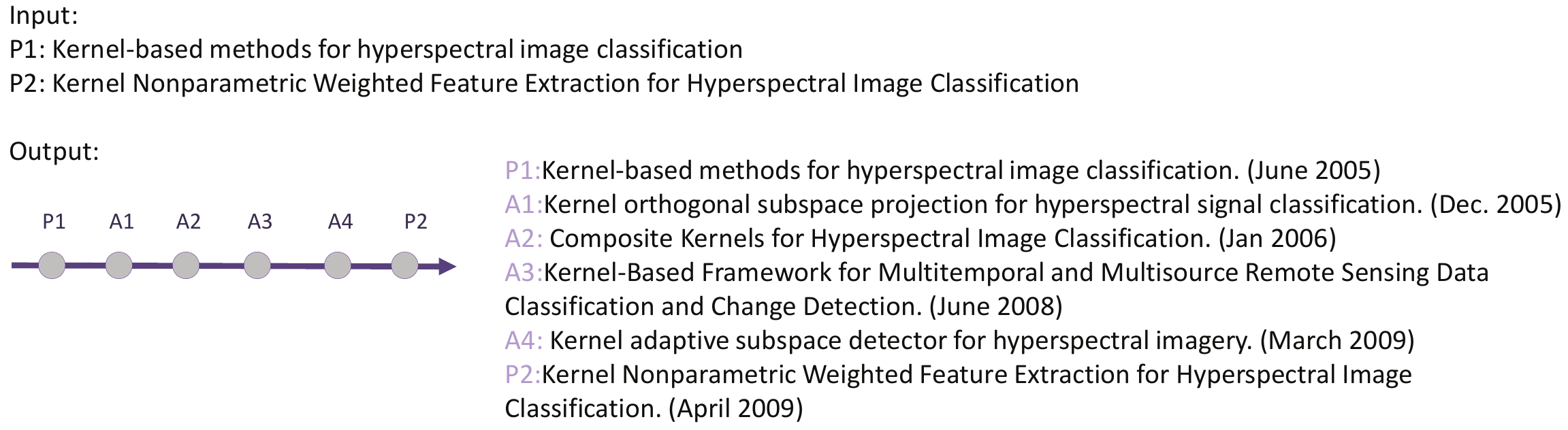}}

\caption{PEG of two papers retrieval focusing on hyperspectral imagery classification.}\label{fig:PEG of two papers retrieval classification}
\end{figure}
\begin{figure}[t]
  \centering

     {\includegraphics[width=1\linewidth]{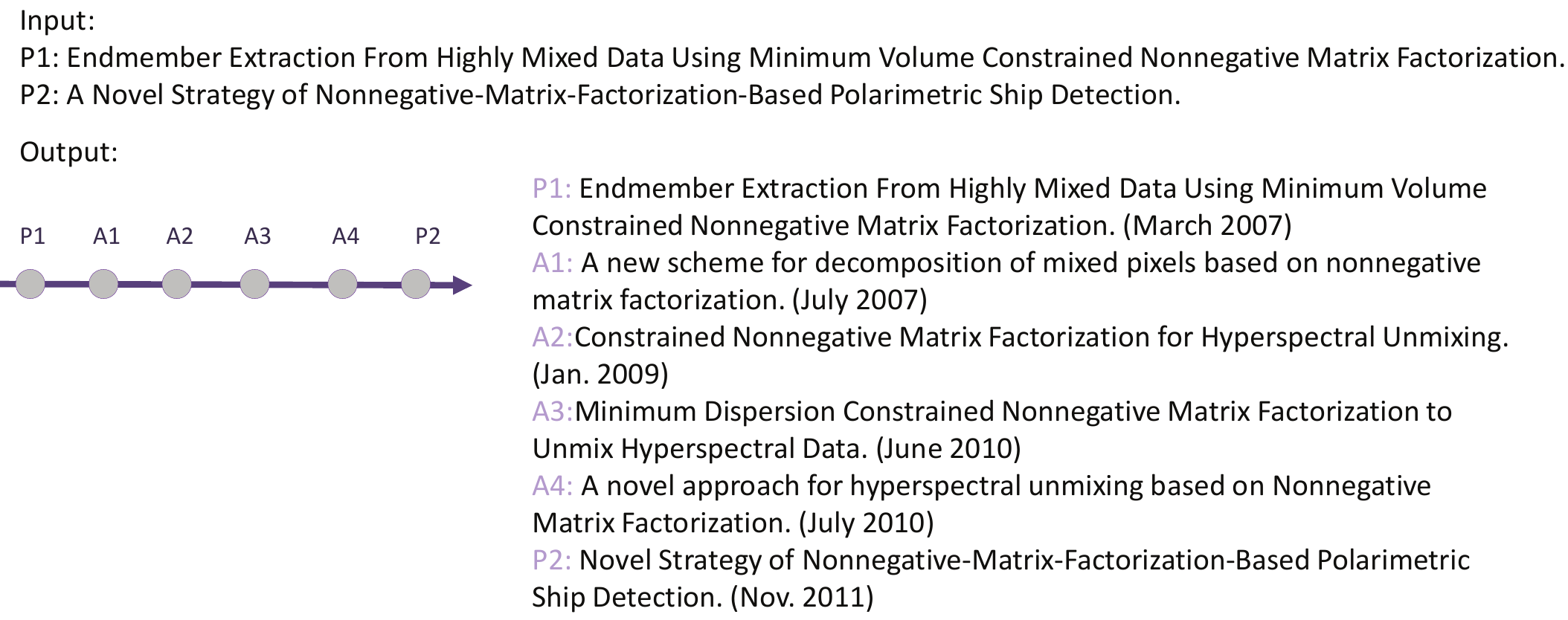}}

\caption{PEG of the two-paper retrieval regarding unmixing and detection using NMF.}\label{fig:PEG of two papers retrieval unmixing}
\end{figure}
\subsection{Search by keyword}
\begin{figure}[t]
  \centering

     {\includegraphics[width=1\linewidth]{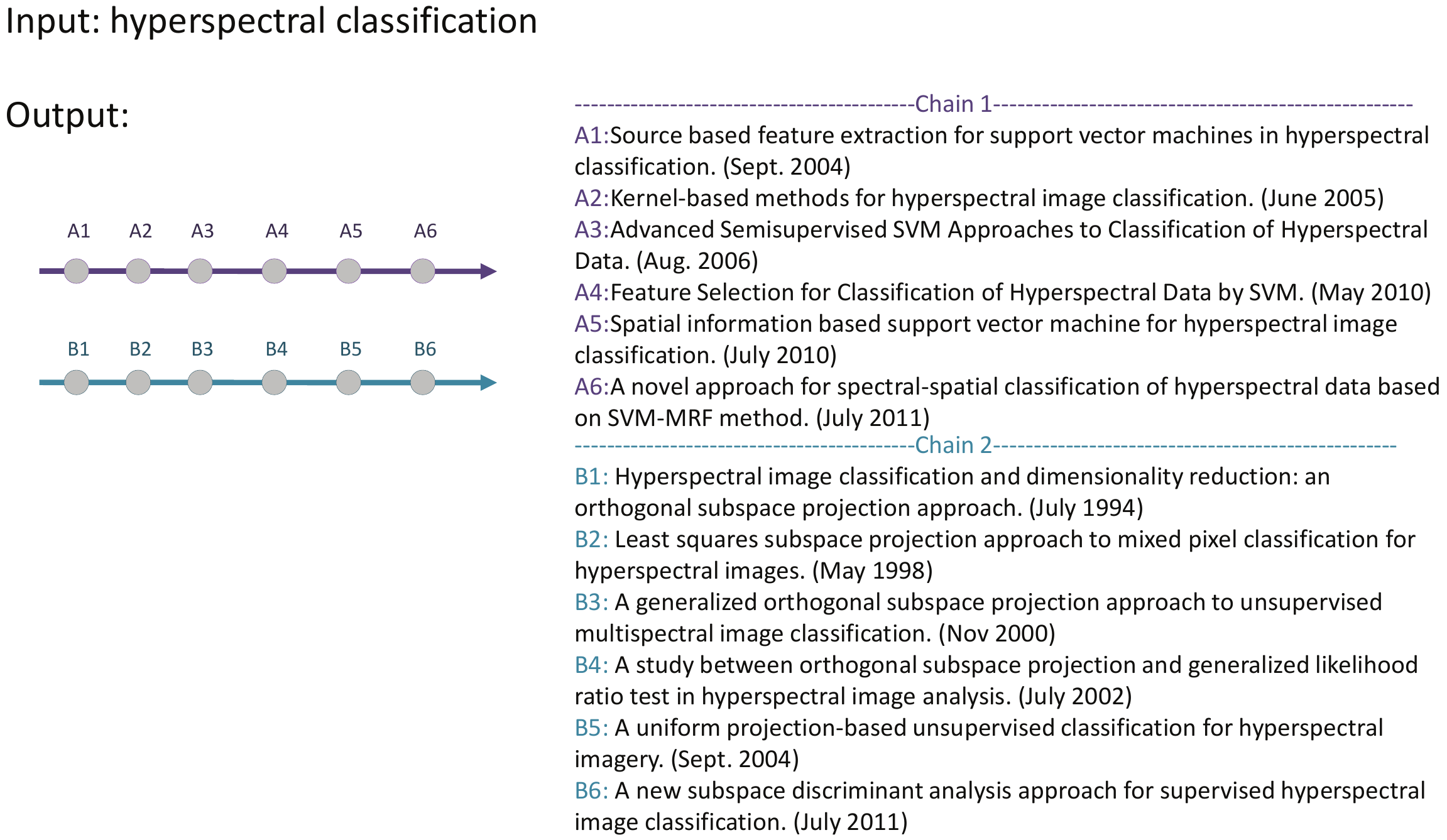}}

\caption{PEG for ``hyperspectral classification'' retrieval.}\label{fig:PEG of hyperspectralclassification}
\end{figure}
\begin{figure}[t]
  \centering

     {\includegraphics[width=1\linewidth]{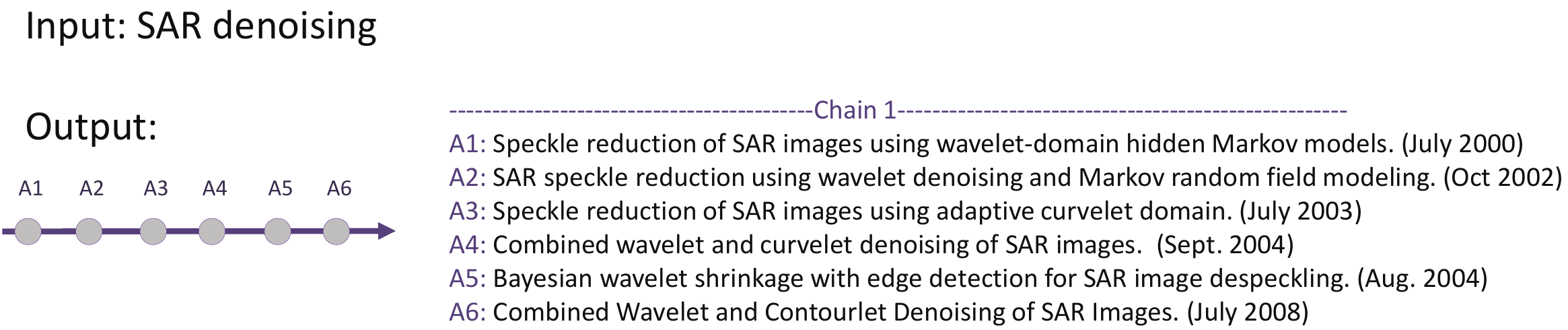}}

\caption{PEG for ``SAR denoising'' retrieval.}\label{fig:PEG of sardenoise}
\end{figure}
For a keyword search, our system chose 100 papers which were most relevant to the keyword as candidates in constructing a PEG.
Fig.~\ref{fig:PEG of hyperspectralclassification} shows the PEG generated for query-words ``hyperspectral classification.''
The length of the evolution chain is set to 6.
Since the candidate papers are from community 11 and community 14,
the PEG consists of two evolution chains extracted from the two communities.
Both chains of papers solve the problem of hyperspectral imagery classification.
Chain 1 follows the technique of classifying hyperspectral pixels using SVM, a classifier that is well known for its superior performance on small data sets with high-dimensional features.
Chain 2 focuses on subspace method-based classification methods, where the high dimensional data are first projected to a low dimensional subspace before classification.

Fig.~\ref{fig:PEG of sardenoise} shows the PEG generated for the keyword ``SAR denoising.''
The evolution chain consists of a set of papers on denoising Synthetic-aperture radar (SAR) images
in the wavelet and curvelet domain.
\subsection{Incorporate user preference}
Our system is able to utilize the three relations between papers to generate PEGs from different views.
We integrate user preferences into our framework.
At the soft-clustering stage, users are allowed to choose different weights $(w_1, w_2, w_3)$ for the three relations according to their needs.
When no user preference is provided, we set $w_1=0.33$, $w_2=0.33$, and $w_3=0.33$ .

When a large weight is chosen on the ``Content'' relation, the clustering approach generates communities based mainly on the content similarity of papers.
As a result, papers in an extracted chain are likely to have a high similarity in content.
Fig.~\ref{fig:content} shows the PEG of the query-paper, ``Hyperspectral Subspace Identification'' with $w_1=0.2$, $w_2=0.6$, and $w_3=0.2$.
It shows that all papers in the chain focus on the subspace-based method  in hyperspectral imagery.

The citation relation reveals the correlation between papers if they do not have a high similarity in content.
If users are interested in discovering papers that have a citation relationship, he/she can choose a large weight on the ``Citation'' relation.
In this case, papers having citation relationships are likely to be clustered into the same community.
As a result, papers in an evolution chain not only share the same topic but also have citation relation along the chain.
Fig.~\ref{fig:citation} shows the PEG of the query-paper, ``Hyperspectral Subspace Identification'' with $w_1=0.6$, $w_2=0.2$, and $w_3=0.2$.
In the evolution chain, both $C3$ and $C4$ cite $P$ and $C4$ cites $C3$.

In the case of a large weight for the ``Author'' relation, our clustering approach is inclined to generate clusters based on the authorship information, i.e., papers published by the same authors are tend to be clustered into communities.
As a result, chains extracted from such communities are likely to consist of papers sharing the same authors.
Fig.~\ref{fig:author} shows the PEG for the query-paper ``Hyperspectral Subspace Identification'' with $w_1=0.2$, $w_2=0.2$, and $w_3=0.6$, where the first two papers share two same authors and the last two papers share a common author.
\begin{figure}[t]
  \centering
      \subfigure[PEG emphasizes  content coherence.]{\label{fig:content}
      \includegraphics[width=0.45\linewidth]{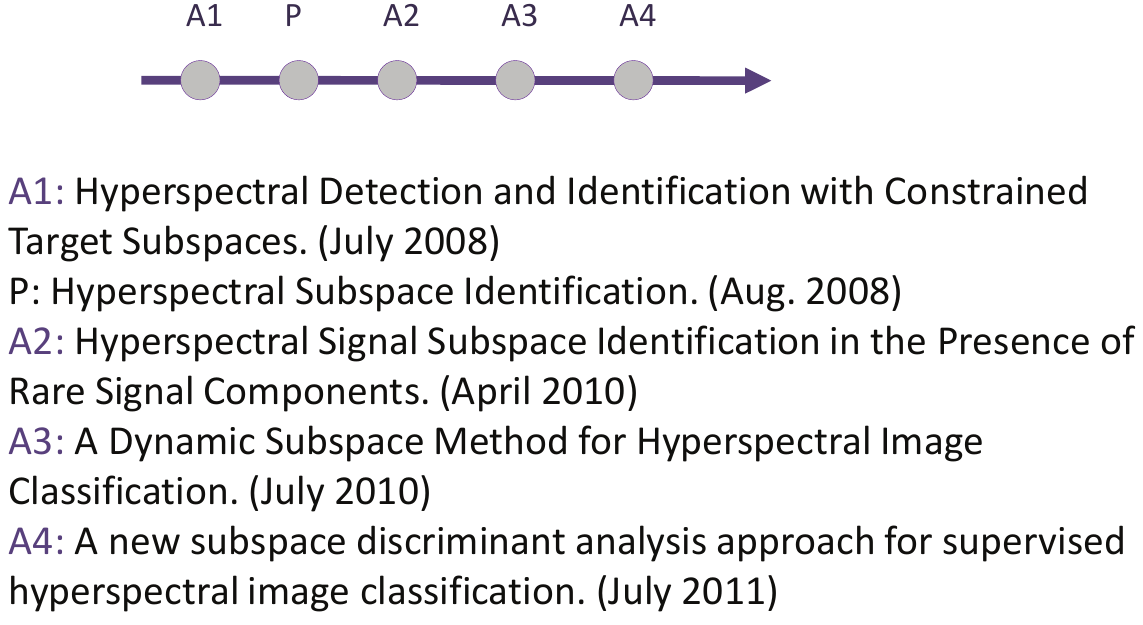}}
      \subfigure[PEG emphasizes citation coherence.]{\label{fig:citation}
      \includegraphics[width=0.45\linewidth]{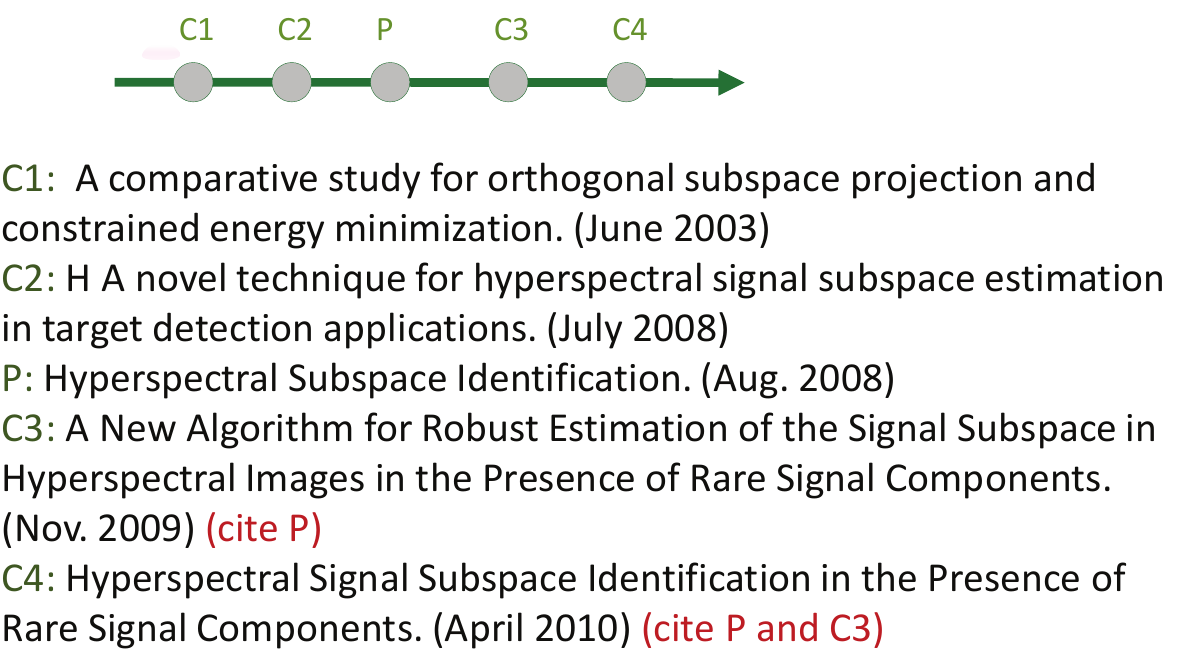}}
      \\
      \subfigure[PEG emphasizes author coherence.]{\label{fig:author}
      \includegraphics[width=0.9\linewidth]{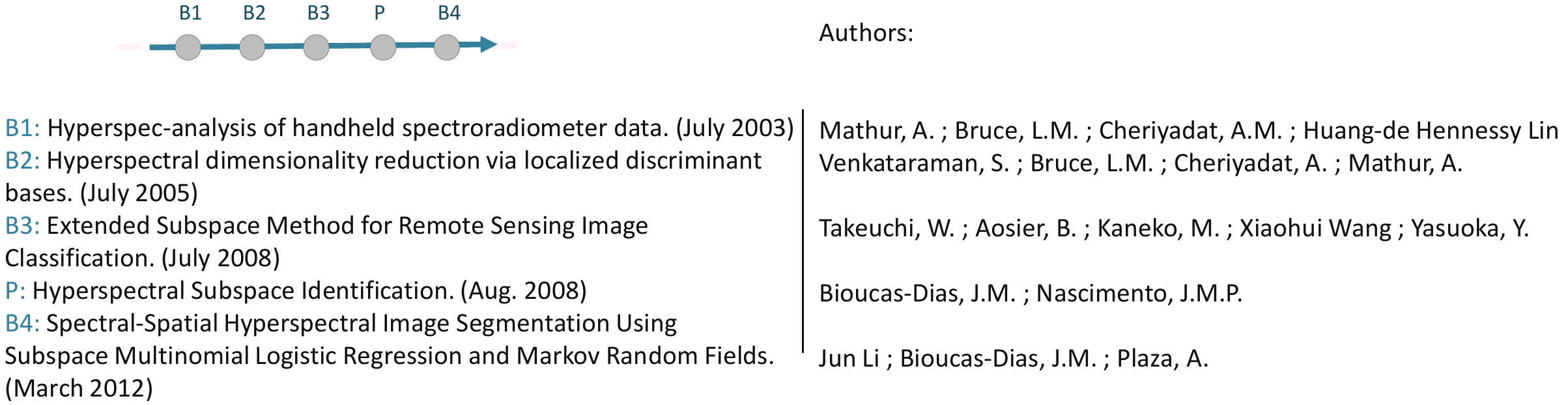}}
\caption{PEGs from different views.}\label{fig:Word influence in paper transition}
\end{figure}
\section{Evaluation}
\label{Evaluation}
A common method to evaluate the performance of IR systems is to test the algorithms on labeled datasets and calculate some standard metrics such as the retrieval precision.
However, these datasets are designed for list-output models and thus they are not suitable for evaluating our structural retrieval system.
As a result, we evaluated our algorithm on real-world data sets and tested three metrics:
1) Accuracy, which  shows how strong the articles retrieved in PEG are relevant to
the query;
2) Coherence, which shows the topic coherence of an evolution chain.
We measured the coherence by the definition given in Equation~\ref{finalobjective} and also invited some domain experts to evaluate the topic coherence of the results.
3) Helpfulness, where we conducted a user study to see how PEG can help beginners understand a new academic article as well as get the big picture on an unfamiliar research domain.

Because the proposed method returns structural retrieval results, it can not be compared directly with most of the other retrieval systems that return isolated results.
In this paper, we designed  a way to compare our system with Google Scholar, IEEE Xplore, and Web of Science by manually constructing evolution chains from the retrieval results of the compared systems.
\subsection{Accuracy}
In this study, we evaluated the accuracy of the PEG, i.e., whether the articles in the PEG are relevant to the query.
To this end, we compared the retrieval results of our system with the results of Google Scholar, IEEE Xplore and Web of Science to see how our retrieval results are ranked in other systems.
Our assumption is that if the articles in the PEG are at the top of the returned list in the other systems, then it indicates that the accuracy of the PEG is high.
We created two tasks corresponding to two keyword queries ``SAR denoising'' and ``hyperspectral classification.''
For a fair comparison, the search region in the compared systems was set to TGRS and IGARSS.
Fig.~\ref{fig:ROCclassification} shows the comparison of results of the query ``hyperspectral classification''.
Fig.~\ref{fig:ROCunmixing} shows the comparison of results of the query ``SAR denoising''.
The data in the compared systems was collected on March 14, 2015.
The horizontal axis represents the rank index of the papers from PEG in the compared systems.
The vertical axis is the proportion of papers covered in the PEG.
For example, the green line in Fig.~\ref{fig:ROCclassification} indicates that 60 percent of the papers in the PEG are ranked before 50 in Google Scholar.
The comparison shows that a large proportion of papers in PEG ranked on top of the retrieval list of the other systems.
Note that since our goal is to find out those articles that are most topically coherent, including the most influential/relevant papers is not the main, desired property of PEG.
\begin{figure}[t]
  \centering
\subfigure[Comparison for ``hyperspectral classification'']
{\label{fig:ROCclassification}\includegraphics[width=0.7\linewidth]{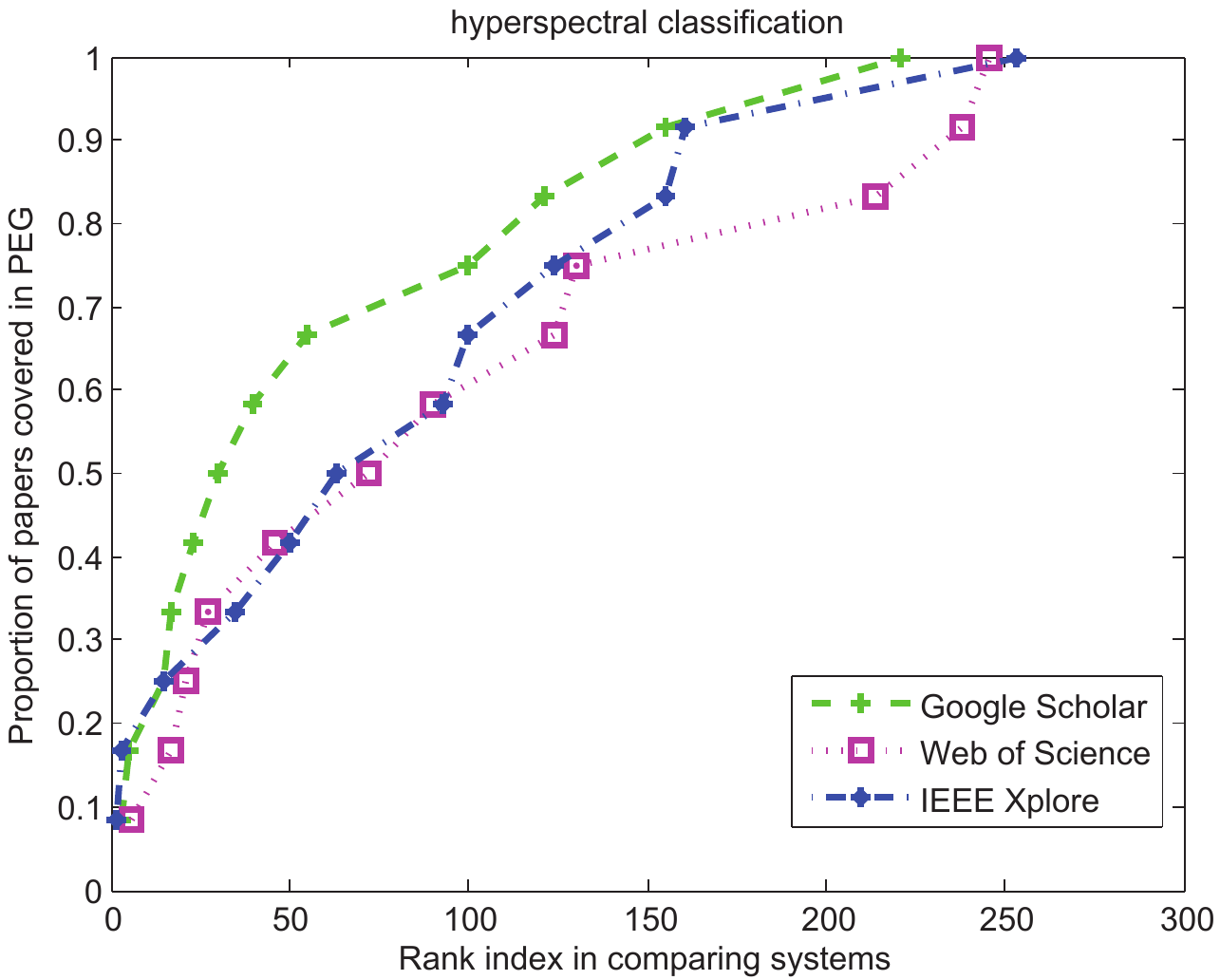}}
\subfigure[Comparison for ``hyperspectral unmixing'']
{\label{fig:ROCunmixing}\includegraphics[width=0.7\linewidth]{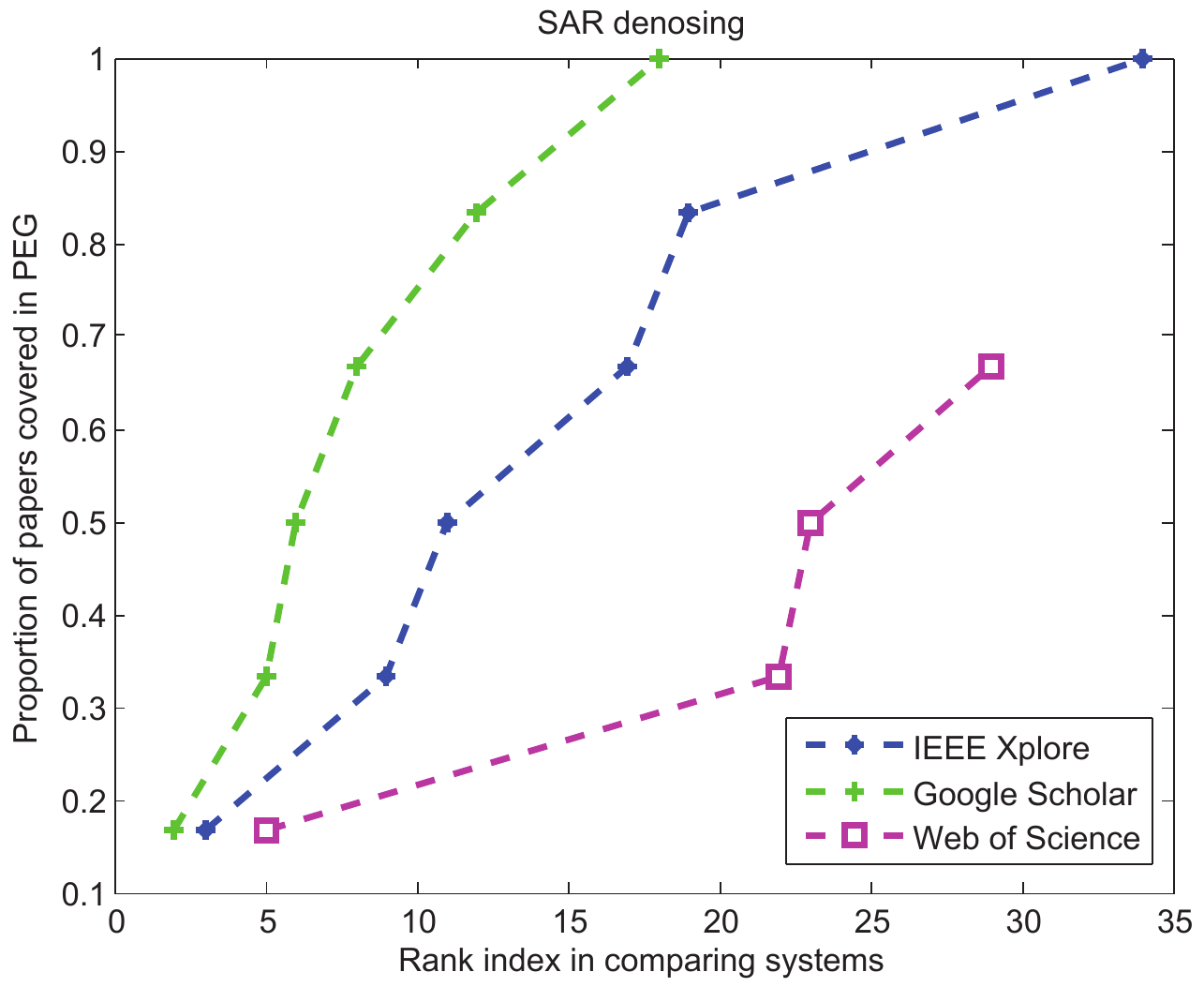}}
\caption{Proportion of papers in PEG covered by different systems.}\label{fig:ROC}
\end{figure}
\subsection{Coherence}
\label{sectioncoherence}
In this section, our goal is to evaluate the topic coherence of the PEG. This is done by measuring the topic coherence defined in Equation~\ref{finalobjective}.
We also invited several domain experts to manually score the coherence of the retrieval results.
We compared our method with the retrieval results from Google Scholar, IEEE Xplore and Web of Science.

Since the retrieval results of the compared systems are isolated, they can not be compared with PEG directly.
For a fair comparison, we manually constructed evolution chains from the retrieved papers of the compared systems.
Again, the papers in the compared systems were restricted to publication in TGRS and IGARSS.
The comparison was done under three types of query.
The length of the evolution chain was set to $6$ in all experiments.
For different types of queries, we constructed chains from the compared systems according to the following steps:

For a single-paper query, we first selected the top $5$ papers in the returned list from the compared systems.
Then the selected papers together with the query paper were arranged in chronological order to form an evolution chain of length $6$.
When the PEG contained more than one chain, $5$ more papers were selected from the compared systems to form each additional chain.

For a two-paper query($p_s$ or $p_t$), we constructed a chain from the systems in two steps.
First, for each of the two queries, we selected its two most relevant papers whose publication time was between that of $p_s$ and $p_t$.
Then the four selected results together with $p_s$ and $p_t$ were arranged in chronological order to form an evolution chain.

For a keyword query, we first selected the top $6$ papers from the compared systems to form an evolution chain.
Six more papers were selected for each additional chain in the PEG.
Then the papers were arranged in chronological order to form the same number of chains as in the PEG.

We created ten tasks for each type of query and calculated the topic coherence as defined in Equation~\ref{finalobjective}.
Fig.~\ref{fig:Comparison of topic coherence} shows the average coherence of the tasks for different systems.
The result demonstrates that the evolution chains in PEG are much more topically cohesive than the other systems' results.
\begin{figure}[t]
  \centering
     {\includegraphics[width=0.8\linewidth]{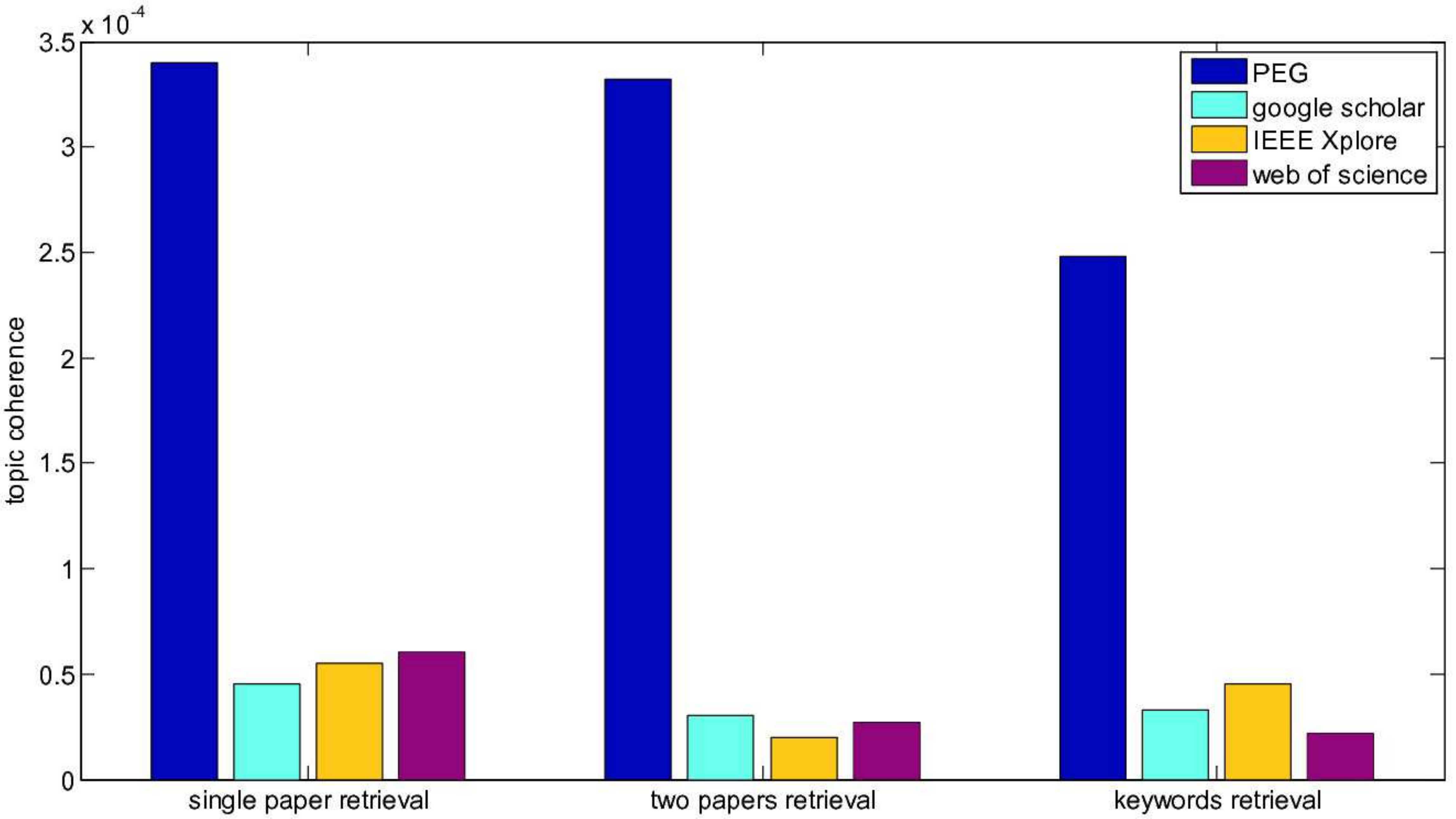}}
\caption{Comparison of topic coherence.}\label{fig:Comparison of topic coherence}
\end{figure}

We also evaluated the coherence by asking a group of domain experts to score the chains generated by different systems.
Since the PEG is unique in structure (chains intersects with each other), we were unable to do a double-blind comparison study in its original form.
To deal with this problem, we separated the combined chains in PEG into independent chains so that the experts could not differentiate between the different systems.
The expert group was composed of 6 experts, including four teachers and two Ph.D. students.
Each expert had at least two years of research experience in the remote sensing domain.
We created five tasks for each type of query in the study.
The experts were asked to score the correlation between adjacent papers with 0-5 points (0 points means not related at all).
Since a high correlation between the adjacent papers does not guarantee a coherent topic along the whole chain, we also asked the experts to grade the topic coherence of the overall chain with 0-5 points.
One of the grading tables is shown in Fig.~\ref{fig:User study table in the expert group}.
\begin{figure}[t]
  \centering

     {\includegraphics[width=1\linewidth]{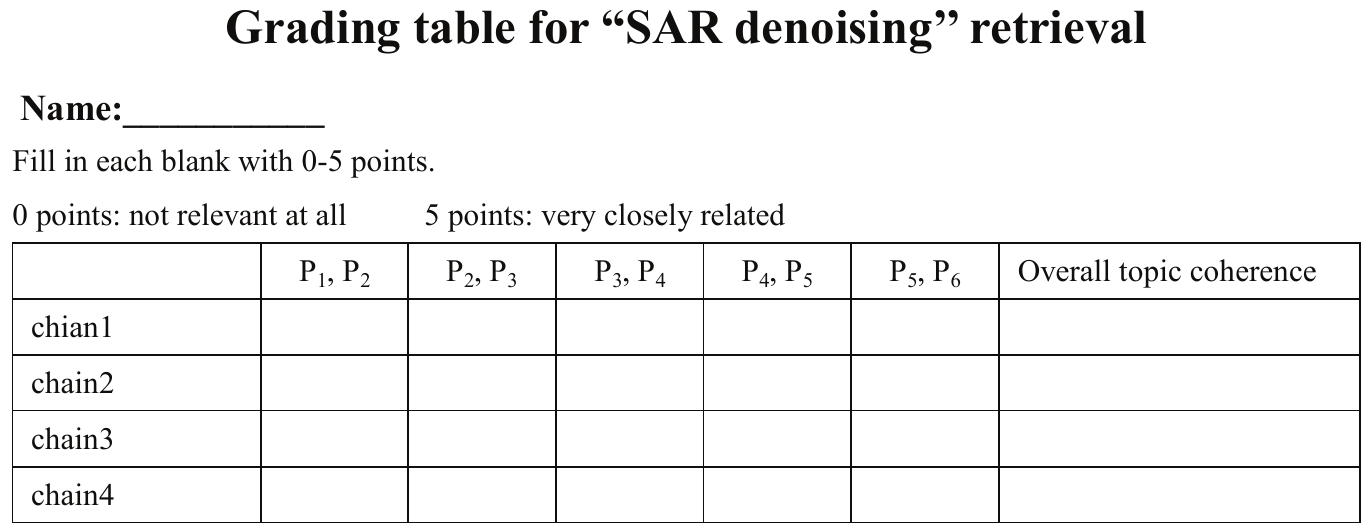}}

\caption{Grading table for ``SAR denoising'' retrieval.}\label{fig:User study table in the expert group}
\end{figure}
The average grading results for all tasks are given in Table~\ref{Average score of the expert group}.
In the study, the experts gave very high marks to the retrieval results presented by PEG.
The results show that the correlations between adjacent papers in PEG are much stronger than in other systems.
In addition, the topic along the overall chain of PEG was
 much more focused  than in the compared systems.
\begin{table}[t]
  \caption{Average coherence score of the expert group.}
  \label{Average score of the expert group}
  \centering
  \begin{tabular}{ccc}
    \hline
    Coherence & Adjacent score & Overall score\\
     \hline
    PEG & $4.1$ & $4.53$\\
    Google Scholar & $2.4$ & $2.45$\\
    Web of Science & $1.64$ & 2.2 \\
    IEEE Xplore & $1.35$ & $1.78$\\
    \hline
  \end{tabular}
\end{table}
\subsection{Helpfulness in digesting new information}
In this section, we analyze whether PEG can help beginners digest new information
when facing an unfamiliar research domain.
To achieve this goal, we designed a user study to see how PEG can help users understand a new paper as well as comprehend the big picture of a new research domain.
We recruited 16 students in our college to do the study.
All of the participants satisfied two conditions: 1. They were able to read academic
articles in the remote sensing domain. 2. They did not know the domain well in advance.
In this study, we designed three tasks for a single-paper query and a keyword query, respectively.
In the single-paper query, we aimed to find out how PEG can help beginners answer specific questions related to the query-paper.
In the keyword search tasks, the users were asked to describe some important concepts and answer specific questions regarding the query-domain.
The queries and questions were designed specifically for beginners by a Ph.D. student who had rich research experience in the remote sensing domain.
Still, we compared the level of knowledge attained by beginners using our prototype versus Google Scholar, IEEE Xplore, and Web of Science.
The approach used to select papers from the compared systems is described in section~\ref{sectioncoherence}.
The students were divided into four groups to score the results from
the different systems.

At the first stage of the single-paper query test,
each participant was asked to answer a questionnaire with questions
regarding the query-paper.
One correct answer added one point.
The scores were recorded to measure the students' pre-knowledge about the paper.
After they finished the questionnaires, students in different groups
were asked to read the query-paper with the
help of a set of retrieved papers from the compared retrieval systems.
They were allowed to modify the questionnaires while reading.
After they finished their final questionnaires,
the improved scores of each student were calculated.
We believe that the higher the improved scores were, the more useful the retrieved
papers were in helping the beginners to digest a paper.

The second user study was to measure the helpfulness of PEG in aiding users to grasp
the general information in unfamiliar domains.
Again, before reading any articles the students were asked to answer a questionnaire regarding the three research domains.
Then one group of students was presented with PEG results generated by keyword search,
while students in the other groups were presented with papers retrieved by the compared
systems.
They were allowed to modify their questionnaires while reading.

We recorded the final scores and calculated the  improved scores for all of the six tasks.
The average improved scores of the four groups are given in Table~\ref{Improved scores of different groups}.
The results demonstrate that PEG performs better in helping beginners  digest the details of a paper and comprehend the
general knowledge within a research domain.
\begin{table}
 \centering
\caption{Improved scores of the different groups.}
 \label{Improved scores of different groups}
  \begin{tabular}{ccc}
    \hline
    Improved score & Search by & Search by \\
     ~&single paper&keyword\\
     \hline
    PEG & $4.83$ & $3.20$\\
    Google Scholar & $3.15$ & $2.04$\\
    Web of Science & $2.58$ & $1.28$ \\
    IEEE Xplore & $2.11$ & $1.67$\\
    \hline
  \end{tabular}
\end{table}

\section{Conclusions and future work}
\label{Conclusions and future work}
In this paper, we presented a method for creating structured paper retrieval results, which we call a PEG.
A PEG explicitly shows the multi-view relationships between the retrieved papers by a combination of a set of evolution chains.
Each chain consists of a sequence of topically cohesive papers;
different chains follow different topics of the query.
Three types of information (``Content'', ``Author'' and ``Citation'') are utilized in our system, to which users are allowed to attribute different weights to generate an evolution graph emphasizing different aspects.
Our system supports keyword search, single-paper searches and two-paper
 searches to satisfy different user requirements.

In the future, we plan to enlarge our dataset and invite more researchers to use and evaluate our system.
In addition, we intend to make use of more types of information
such as a paper's academic influence.
Another important issue is to extend our system to time-varying datasets.


\bibliographystyle{IEEEtran}
\bibliography{IEEEabrv,refs}
%
%

\end{document}